\documentclass{article}

\usepackage{arxiv}

\usepackage[T1]{fontenc}
\usepackage[utf8]{inputenc}
\usepackage{lmodern}
\usepackage{microtype}

\usepackage{amsmath,amsfonts,amssymb}
\usepackage{bbm}

\usepackage{booktabs}
\usepackage{array}
\usepackage{tabularx}
\usepackage{longtable}
\usepackage[table]{xcolor}

\usepackage{graphicx}
\graphicspath{{./images/}}

\usepackage{float}

\usepackage{tikz}
\usetikzlibrary{arrows.meta,positioning,fit,backgrounds}
\usepackage{pgfplots}
\pgfplotsset{compat=1.18}
\pgfdeclarelayer{background}
\pgfsetlayers{background,main}

\usepackage[most]{tcolorbox}
\tcbset{
  colback=gray!5,
  colframe=black!15,
  boxsep=2pt,
  arc=2mm,
  left=3mm,
  right=3mm,
  top=2mm,
  bottom=2mm
}

\usepackage{url}
\usepackage{natbib}
\usepackage{doi}
\usepackage{hyperref}

\newcolumntype{L}[1]{>{\raggedright\arraybackslash}p{#1}}
\newcolumntype{Y}{>{\raggedright\arraybackslash}X}

\newcommand{\edited}[1]{\textcolor{black}{#1}}

\title{Diverse Approaches to Optimal Execution Schedule Generation}

\author{
  Robert de~Witt\thanks{The views, opinions and conclusions expressed here are solely those of the authors and do not necessarily reflect the views or policies of the Bank of America, or any other institution with which the authors are affiliated. No responsibility should be attributed to those institutions . This article has not been reviewed, approved, or endorsed by the authors’ employers or any affiliated organizations}\\
  Imperial College London \\
  Bank of America Securities\\
  London, United Kingdom \\
  \texttt{robert.de-witt23@imperial.ac.uk} \\
  \texttt{robert.de\_witt@bofa.com}
  \And
  Mikko S. Pakkanen \\
  Department of Mathematics\\
  Imperial College London\\
  London, United Kingdom \\
  \texttt{m.pakkanen@imperial.ac.uk}
}
\renewcommand{\shorttitle}{Diverse Approaches to Optimal Execution}

\hypersetup{
  pdftitle={Diverse Approaches to Optimal Execution Schedule Generation},
  pdfsubject={Quantitative Finance, Market Microstructure, Reinforcement Learning},
  pdfauthor={Robert de Witt, Mikko S. Pakkanen},
  pdfkeywords={Optimal Execution, Reinforcement Learning, Market Impact, Transient Impact, Quality-Diversity, MAP-Elites, PPO, Implementation Shortfall, Equities},
}

\begin{document}
\maketitle

\begin{abstract}

We present the first application of MAP-Elites, a quality-diversity algorithm, 
to \edited{trade} execution. Rather than searching for a single optimal policy, 
MAP-Elites generates a diverse portfolio of regime-specialist strategies 
indexed by liquidity and volatility conditions. Individual specialists achieve 
8-10\% performance improvements within their \edited{behavioural} niches, while other 
cells show degradation, suggesting opportunities for ensemble approaches that 
combine improved specialists with the baseline PPO policy. Results indicate 
that quality-diversity methods offer promise for regime-adaptive execution, 
though substantial computational resources per \edited{behavioural} cell may be required 
for robust specialist development across all market conditions.

To ensure experimental integrity, we develop a calibrated Gymnasium environment 
focused on order scheduling rather than tactical placement decisions. The 
simulator features a transient impact model with exponential decay and 
square-root volume scaling, fit to 400+ U.S. equities with $R^2>0.02$ 
out-of-sample. Within this environment, two Proximal Policy Optimization 
architectures---both MLP and CNN feature extractors---demonstrate substantial 
improvements over industry baselines, with the CNN variant achieving 2.13 bps 
arrival slippage versus 5.23 bps for VWAP on 4,900 out-of-sample orders 
(\$21B notional). These results validate both the simulation realism and 
provide strong single-policy baselines for quality-diversity methods.
\end{abstract}


\newpage
\section{Introduction}\label{sec:Introduction}

\begin{figure}[t]
\centering
\begin{tikzpicture}[
    node distance=0.7cm and 0.7cm,
    >=Stealth,
    thick,
    block/.style={rectangle, draw, rounded corners, text width=4.25cm, align=center, minimum height=3cm},
    main_block/.style={block, very thick, text width=6cm},
    background_block/.style={rectangle, draw, rounded corners, fill=gray!10, inner sep=0.75cm}
]

\node (empirical) [block, fill=blue!20] 
    {Empirical Impact Models\\ (e.g., Bouchaud Propagator)\\ \footnotesize Calibrated decay kernel, transient impact, concave scaling};

\node (classical) [block, fill=blue!10, left=of empirical] 
    {Classical Models\\ (e.g., Almgren--Chriss)\\ \footnotesize Mean--variance optimisation, constant volatility, stationary impact};

\node (rl) [block, fill=orange!20, right=of empirical] 
    {Reinforcement Learning Approaches\\ \footnotesize Agent--environment loop, adaptive policies, simulation-trained};

\node (thiswork) [main_block, fill=yellow!30, below=of empirical, yshift=-1.0cm] 
    {This Work:\\Novel RL Optimal Execution \\ \footnotesize PPO, MAP-Elites, High-fidelity Gymnasium environment, calibrated transient impact, realistic order generation, vectorised simulation};

\draw[->] (classical.south) -- (thiswork.north west);
\draw[->] (classical.east) -- (empirical.west);
\draw[->] (empirical.south) -- (thiswork.north);
\draw[->] (empirical.east) -- (rl.west);
\draw[->] (rl.south) -- (thiswork.north east);

\begin{pgfonlayer}{background}
    \node [
        background_block, 
        fit=(classical)(rl)(thiswork), 
        label={[font=\large\bfseries]above:Optimal Execution}
    ] {};
\end{pgfonlayer}

\end{tikzpicture}
\caption{Evolution of optimal execution approaches: from classical models to empirical impact models to reinforcement learning, with this work positioned at the intersection of empirically calibrated models and RL methods.}
\label{fig:oe_evolution}
\end{figure}
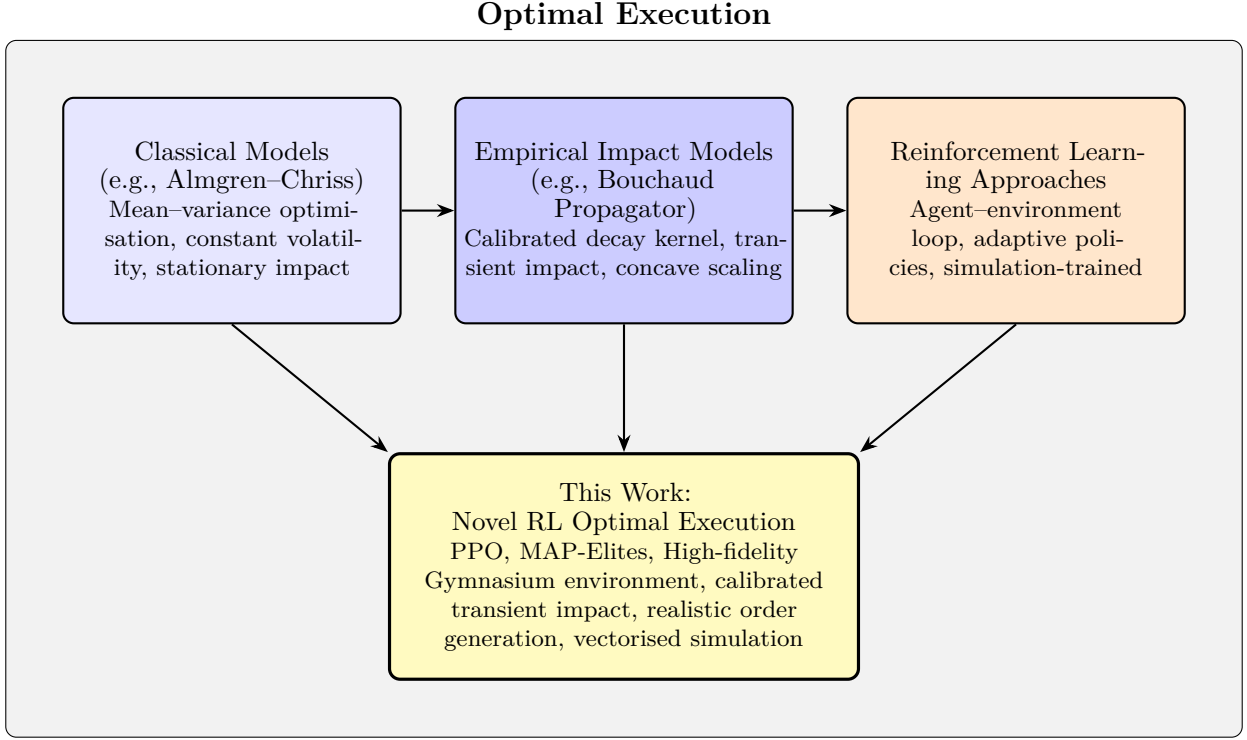

Optimal execution (OE) is a central problem in algorithmic trading, influencing approximately a trillion dollars of daily turnover across global equities and futures markets. It concerns determining how to trade a given order over a predetermined or dynamic horizon while minimising transaction costs relative to a benchmark, typically the \emph{arrival price}—the mid-quote at the time the order is initiated. This performance is often expressed as \emph{implementation shortfall} \citep{Perold1988IS}, the difference between the value of an ideally priced portfolio and the actual cost of implementing it through trading. Transaction costs arise from explicit sources (e.g., commissions and fees) and implicit sources (e.g., market impact and slippage). The execution challenge is compounded by the stochastic nature of prices, variable liquidity conditions, and the trade-off between market impact and timing risk \citep{AlmgrenChriss2001JOR}.

\edited{The evolution of optimal execution approaches is outlined in Figure~\ref{fig:oe_evolution}.} Traditional approaches, such as the seminal Almgren-Chriss framework, cast the execution problem as a mean-variance optimisation in which market impact is modelled as an additive cost and risk is penalised via price variance. While yielding tractable closed-form schedules (e.g., linear, front-loaded, or back-loaded trajectories), these models assume constant volatility, stationary impact functions, and exogenous order flow \citep{AlmgrenChriss2001JOR, ObizhaevaWang2013}. Empirical studies show that market impact scales concavely with order size and decays transiently over time \citep{Bouchaud2010PriceImpact, bouchaud2018propagator}, motivating richer dynamic models.

In recent years, alongside other data-driven approaches, reinforcement learning (RL) has emerged as a promising alternative for OE \citep{nevmyvaka2006reinforcement, hendricks2014reinforcement}. RL agents can learn adaptive policies that respond to evolving market states without assuming explicit parametric forms for price dynamics or impact decay. The agent-environment loop in Gymnasium \edited{\citep{towers2024gymnasium}} offers a natural abstraction: the agent observes the current market state (e.g., prices, volumes, volatility, time remaining, inventory) and outputs an action representing a trade size or participation rate. The environment then simulates execution, applies market impact, updates the state, and returns a reward signal tied to execution performance. This process enables the agent to learn policies through simulation without incurring the cost and risk of live experimentation; once deployed, such policies can be further adapted using live trading outcomes.

However, much of the existing RL literature for OE suffers from limited realism in backtesting. Many implementations use oversimplified price dynamics (e.g., geometric Brownian motion) or neglect empirically calibrated market impact, while others focus on highly granular limit order book (LOB) simulations that require many structural assumptions and can diverge from practical execution workflows. This \edited{paper} takes a middle path, avoiding both coarse-grained price-only models and overly complex LOB simulations. It builds a high-fidelity Gymnasium-based back-testing environment calibrated to one year of historical minute-bar data for hundreds of US equities. The environment integrates a transient market impact model fit via cross-validation, realistic order-arrival processes, and configurable state/reward designs, enabling a robust evaluation of RL and baseline strategies.

We utilise this environment to train RL policies designed to decide how much to trade at any point in an order's lifespan. The generated policy is not concerned with what limit price, type, or venue to use, but rather the schedule of quantities to trade, given a reward-driven objective. With the right reward structure and cost function, this approach could, in principle, generalise beyond \edited{volume-weighted average price (VWAP)} to improve upon present-day execution schedules such as \edited{time-weighted average price (TWAP)}, \edited{percentage of volume (POV)}, implementation shortfall, or liquidity-seeking strategies. We focus on an optimised VWAP structure which aims to reduce slippage to the order arrival price by allowing some discretion to a base VWAP schedule. We choose this as an initial approach as there are immediate large-scale applications for an optimally performing scheduler. This structure greatly simplifies the action space of the RL algorithm while opening up flexibility for a highly informative state space to act on, increasing the likelihood of superior decisions can be learned in simulation.

A key limitation of existing RL approaches to execution is that they optimise 
for a single policy maximizing average performance across all market conditions. 
Quality-diversity (QD) algorithms \edited{\citep{Chatzilygeroudis2021}}, developed originally for adaptive robotics, 
offer an alternative paradigm. Rather than searching for a single optimum, QD 
methods such as MAP-Elites \citep{mouret2015mapelites} generate portfolios of 
high-performing policies, each specialised for different \edited{behavioural} niches 
defined by descriptor features. In robotics, these descriptors might characterise 
terrain type or gait stability; in optimal execution, natural descriptors include 
market regime (volatility, liquidity), order urgency, or directional momentum. 
We apply MAP-Elites using liquidity and volatility as \edited{behavioural} descriptors to 
generate regime-specialist execution policies. To our knowledge, this is the 
first application of quality-diversity methods to the optimal execution problem.

The contributions of this work are:
\begin{enumerate}
\item \textbf{First application of quality-diversity methods to financial 
execution.} We apply MAP-Elites to generate portfolios of regime-specialist 
execution policies indexed by liquidity-volatility descriptors. While 
individual specialists achieve 8-10\% improvements in specific niches, 
results reveal challenges in regime classification and training data density 
that motivate future research. To our knowledge, this is the first exploration 
of quality-diversity algorithms for trading.

\item \textbf{Validation of RL under empirically calibrated transient impact.} 
Using a propagator model with exponential decay and square-root scaling 
($R^2>0.02$ out-of-sample), we demonstrate PPO-CNN achieves 59\% lower arrival 
slippage than VWAP (2.13 vs 5.23 bps) on \$21B test set. This establishes 
both a strong baseline for quality-diversity methods and validates that RL 
can exceed industry benchmarks under realistic impact dynamics.

\item \textbf{GEO: Open-source environment enabling reproducible execution 
research.} We \edited{plan to} release a calibrated Gymnasium simulator with realistic order 
generation, minute-bar execution, and transient impact modeling. This 
infrastructure enables fair comparison of execution algorithms and supports 
future quality-diversity research.
\end{enumerate}

The remainder of this paper is structured as follows. Section~\ref{sec:methods} introduces the problem formulation and proposed solutions. We begin with the optimal execution set-up, including the formal problem statement, the transient impact model with its calibration, and the construction of raw and derived features. We then introduce reinforcement learning, translating OE into the RL framework and outlining the order generation process. Next, we present the RL methods explored, covering architecture variations (MLP and CNN) of Proximal Policy Optimisation (PPO) and Quality-Diversity approaches (MAP-Elites) and industry-standard execution strategies. Finally, we describe how these components fit into the simulation environment design, including Gymnasium integration and the execution simulator with impact. Section~\ref{sec:results} reports the impact model calibration results and compares the performance of the RL agents against standard benchmarks. Section~\ref{sec:discussion} concludes with implications of our findings, limitations of the framework, and potential directions for future research.

\section{Methods}\label{sec:methods}

\subsection{Optimal Execution Problem Set-up}\label{subsec:probform}

\subsubsection{Fundamental OE problem formulation}

We consider the execution of a single parent order of \edited{size} $Q_0$ shares over a fixed horizon of $H$ discrete time steps, each corresponding to one minute of market time. The order size $Q_0$, horizon $H$, and stock $S_0$ are specified by the portfolio manager. While one could also consider continuous or event-driven time-steps, for simplicity we work with minute-binned outcomes.

The objective is to minimise the \emph{implementation shortfall} (IS) relative to the mid-price $(p_{\text{bid}}+p_{\text{ask}})/2$ at the start of the order, $p_0$:
\begin{equation*}
\mathrm{IS} = \mathrm{side} \times \left( \frac{\sum_{t=0}^{H-1} p^{\mathrm{fill}}_t \cdot |q_t|}{Q_0} - p_0 \right),
\end{equation*}
where $\mathrm{side} \in \{+1,-1\}$ indicates a buy or sell order, and $q_t$ is the signed quantity traded at time $t$ at the price $p^{\text{fill}}_t$. This expression is equivalent to the realised VWAP of the executed order minus the benchmark price, scaled by trade direction.

The realised fill price, in reality, is the volume weighted average price (VWAP) of the shares traded in the market. In simulation, \edited{$p^{\mathrm{fill}}_t$} incorporates both the prevailing historical market VWAP and the total price impact $I_t$ (immediate plus propagated) generated by the trade. Here $I_t$ is modelled through the transient propagator framework \citep{Bouchaud2010PriceImpact, bouchaud2018propagator}, detailed in Section~\ref{subsec:propagator}. Incorporating such impact-adjusted fills is one of the key differentiators of this work to ensure realism when training our models.

Since the arrival price is only a single point in time benchmark, to better measure our execution efficiency within the horizon $H$, we also consider a second benchmark for robustness: \edited{slippage relative to market VWAP}
\begin{equation*}
P^{\mathrm{VWAP}} = \frac{\sum_{t=0}^{H-1} v_t \, p_t}{\sum_{t=0}^{H-1} v_t},
\end{equation*}
where $v_t$ is traded volume and $p_t$ is the filled price at time $t$.

\subsubsection{Propagator Model (Transient Impact Model)}\label{subsec:propagator}

As with all reinforcement learning policy optimisations, the realism of the simulation environment directly impacts the quality of the learned action policy. If the simulation dynamics diverge materially from the ``physical laws'' governing markets, then the resulting strategies will be suboptimal when deployed. While exchange microstructure rules can be simulated with reasonable fidelity, modelling the impact of incremental orders on market prices \edited{at one-minute granularity} is far more challenging. Fortunately, thanks to the work of \citet{Bouchaud2010PriceImpact, bouchaud2018propagator, ObizhaevaWang2013, gatheral2012transient}, the \emph{propagator model} provides a tractable framework to capture transient market impact. This formulation allows the impact of each executed trade to propagate forward in time with a decaying influence on prices. Given our \edited{one-minute granularity}, where there may be intermittent gaps, we find the transient impact model with an \emph{exponential decay kernel} to be the most suitable for simulation.

\paragraph{General formulation.}
Let $\epsilon_t \in \{+1,-1\}$ denote the trade sign (buy or sell), $q_t$ 
the agent's traded quantity, and $V_t$ the market volume at time $t$. The 
return at time $t$ is modelled as
\begin{equation}\label{eq:propagator_full}
r_t = \sum_{\ell=1}^{L} G(\ell) \, f(q_{t-\ell}, V_{t-\ell}) \, \epsilon_{t-\ell} + \eta_t,
\end{equation}
where $G(\ell)$ is the propagator kernel describing how impact at lag $\ell$ 
decays over time, $f(q, V)$ is the instantaneous impact function, and $\eta_t$ 
is exogenous noise.

The cumulative transient impact $I_t$, which shifts the fill prices in the execution simulator (cf. Section~\ref{subsec:probform}), is then
\begin{equation*}
I_t = \sum_{\ell=1}^{L} G(\ell)\, f(v_{t-\ell})\, \epsilon_{t-\ell}.
\end{equation*}

\paragraph{Instantaneous impact.}
The instantaneous impact function scales as a power law in the participation rate:
\begin{equation*}
f(q, V) = \gamma \left(\frac{q}{V}\right)^\beta, \quad \beta \in (0,1),
\end{equation*}
where $\gamma$ is a stock- and regime-dependent scale factor, and $\beta$ 
typically lies between 0.4 and 0.7 for equities.
 This concavity captures the empirically observed ``square-root law'' of market impact. At the shortest time scales, however, the impact function is often observed to be closer to linear. For example, \citet{Cont2010impact} show that short-horizon price changes are linearly related to order flow imbalance, \citet{toth2011anomalous} find an additive linear response kernel across traders, and \citet{Bucci2019crossover} document a crossover regime between linear and square-root impact. In Section~\ref{sec:results}, we empirically compare both functional forms at the one-minute horizon, finding that the square-root law provides superior explanatory power ($R^2$) in our dataset.

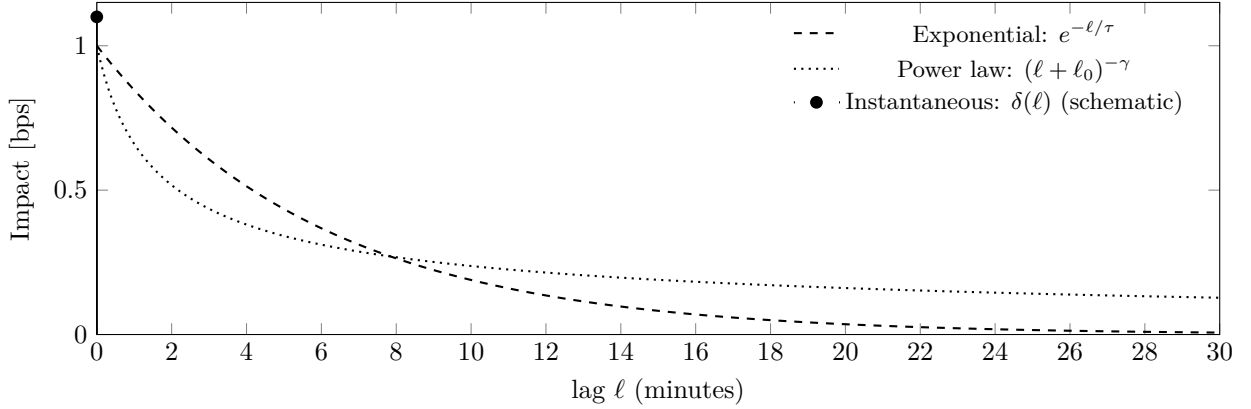
\begin{figure}[t]
\centering
\begin{tikzpicture}
\begin{axis}[
  width=\linewidth, height=6cm,
  xmin=0, xmax=30, ymin=0, ymax=1.15,
  xlabel={lag $\ell$ (minutes)}, ylabel={Impact [bps]},
  legend style={draw=none, fill=none, font=\small, at={(0.98,0.98)}, anchor=north east},
  tick label style={/pgf/number format/fixed},
  domain=0:30, samples=300
]

\addplot[thick, dashed] {exp(-x/6)};
\addlegendentry{Exponential: $e^{-\ell/\tau}$}

\addplot[thick, dotted] {(x+1)^(-0.6)};
\addlegendentry{Power law: $(\ell+\ell_0)^{-\gamma}$}

\addplot[ycomb, mark=*, mark options={solid}, thick] coordinates {(0,1.1)};
\addlegendentry{Instantaneous: $\delta(\ell)$ (schematic)}

\end{axis}
\end{tikzpicture}
\caption{Illustrative transient impact kernels in basis points. The instantaneous kernel is shown schematically as a unit impulse; exponential and power-law forms capture transient, decaying impact with different memory.}
\label{fig:kernel_shapes}
\end{figure}

\
\paragraph{Propagator kernel.}
\edited{As illustrated in Figure~\ref{fig:kernel_shapes}, t}he choice of kernel $G(\ell)$ determines how quickly past trades lose their influence on current prices. Empirical studies show that impact is neither permanent (constant $G(\ell)$) nor purely instantaneous (delta kernel), but decays gradually over time. Several functional forms have been proposed, including power-law kernels \citep{Bouchaud2010PriceImpact} and stretched exponentials \citep{mastromatteo2014agent}.

In this work, we adopt the exponential kernel
\begin{equation*}
G(\ell) = G_0 \, e^{-\ell / \tau},
\end{equation*}
where $G_0$ is the immediate impact coefficient, $\ell$ is the time since the last trade and $\tau$ is the characteristic decay horizon in minutes. This form balances tractability with fidelity: it ensures that impact is transient, avoids long-memory tails that can destabilise calibration on finite samples, and is consistent with empirical fits of minute-bar data. The exponential kernel also integrates naturally with reinforcement learning by ensuring a well-behaved, Markovian state evolution.

\paragraph{Summary.}
Combining the exponential kernel with the power-law impact function yields 
the full transient impact model:
\begin{equation*}
I_t = \sum_{\ell=1}^{L} G_0 \, e^{-\ell / \tau} \, \gamma \left(\frac{q_{t-\ell}}{V_{t-\ell}}\right)^\beta \, \epsilon_{t-\ell},
\end{equation*}
which adjusts fill prices according to:
\begin{equation*}
p^{\mathrm{fill}}_t = p^{\mathrm{VWAP}}_t \, \bigl(1 + \mathrm{side} \cdot I_t \bigr).
\end{equation*}

This formulation arises naturally from resilience models 
\citep{ObizhaevaWang2013}, where impact decays as $\dot{I}(t) = -\tfrac{1}{\tau} I(t) + \kappa \dot{Q}(t)$. 
\citet{gatheral2012transient} showed that admissible (non-manipulable) kernels 
must be completely monotone—i.e., mixtures of exponentials—further justifying 
this choice. While Bouchaud's propagator framework \citep{bouchaud2018propagator} 
often employs power-law kernels, the exponential form offers a tractable 
Markovian approximation that calibrates well at minute-bar horizons.
\subsection{Reinforcement Learning Models}\label{sec:rlmethods}

We investigate several reinforcement learning (RL) approaches as candidates for improving execution performance beyond traditional benchmark methods such as TWAP, VWAP, and POV. Our analysis begins with variations of Proximal Policy Optimisation (PPO) \citep{schulman2017ppo}, a widely adopted and robust policy-gradient algorithm that has become a standard baseline in sequential decision-making. Building on this foundation, we extend our study to a more exploratory direction: MAP-Elites \citep{mouret2015mapelites}, a quality-diversity algorithm designed to promote behavioural diversity while retaining high-performing strategies. To the best of our knowledge, MAP-Elites has not previously been applied to optimal execution, making its evaluation in this setting a novel contribution of our work.

\subsubsection{RL Fundamentals}\label{subsec:rl_fundamentals}

\begin{figure}[t]
  \centering
  \includegraphics[width=0.5\textwidth]{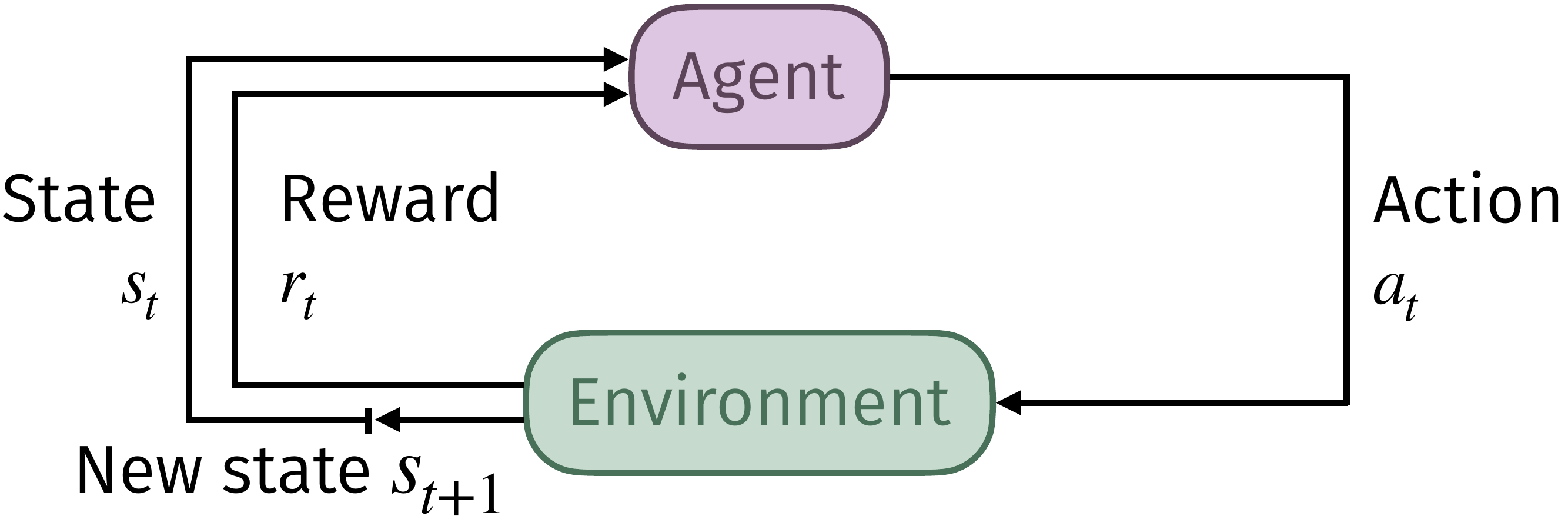}
  \caption{Classic RL Flow diagram\edited{, adapted} from \citet{sutton2018reinforcement}\edited{.}}
  \label{fig:rl_diagram}
\end{figure}

Reinforcement Learning (RL) is a framework for sequential decision-making in which an agent interacts with an environment in order to maximise cumulative reward \citep{sutton2018reinforcement}. \edited{As shown in Figure~\ref{fig:rl_diagram}, a}t each discrete time step $t$, the environment is described by a \emph{state} $s_t \in \mathcal{S}$ that captures the relevant features observable by the agent. In the context of execution, this state might include remaining inventory, elapsed time, spreads, volatility, imbalance, recent trade volumes, prices, or other relevant market information.

The agent selects an \emph{action} $a_t \in \mathcal{A}$, which in execution corresponds to how much of the parent order to trade in the next step (for example, a fraction of the current market volume or a deviation from a baseline schedule). The environment responds by transitioning to a new state $s_{t+1}$ and producing a scalar \emph{reward} $r_t \in \mathbb{R}$, which evaluates the quality of the action taken. In execution, rewards are typically designed as the negative of slippage, transaction cost, or schedule deviation, so that higher returns correspond to better execution quality.

A full sequence of states, actions, and rewards is called a \emph{trajectory},
\begin{equation*}
\tau = (s_0,a_0,r_0,s_1,a_1,r_1,\dots,s_H),
\end{equation*}
with horizon $H$ corresponding to the lifetime of an order. In execution, one trajectory corresponds to a complete order being executed from start to finish.

The objective of RL is to maximise the expected \emph{return} $G_t$, defined as the discounted sum of future rewards:
\begin{equation*}
G_t = \sum_{k=0}^{\infty} \gamma^k r_{t+k},
\end{equation*}
where the \emph{discount factor} $\gamma \in [0,1]$ determines how much future rewards influence present decisions. In execution, setting $\gamma$ close to 1 ensures that the agent considers the long-term cost of completing an order, while smaller values of $\gamma$ emphasise immediate trading costs.

The expected return under policy $\pi$ can be described using a \emph{value function},
\begin{equation*}
V^\pi(s) = \mathbb{E}[G_t \mid s_t = s, \pi],
\end{equation*}
which measures the expected execution quality from state $s$. More generally, the \emph{action-value function} quantifies the return of taking action $a$ in state $s$ and then following $\pi$ thereafter:
\begin{equation*}
Q^\pi(s,a) = \mathbb{E}[G_t \mid s_t = s, a_t = a, \pi].
\end{equation*}
The \emph{advantage function} refines this by comparing the value of a particular action to the state average:
\begin{equation*}
A^\pi(s,a) = Q^\pi(s,a) - V^\pi(s).
\end{equation*}
In execution, the advantage can be interpreted as whether trading faster or slower than usual at a given state improves performance.

A common challenge in estimating advantages is the high variance of Monte Carlo returns. To mitigate this, \citet{schulman2016gae} introduced \emph{Generalised Advantage Estimation (GAE)}, which mixes $n$-step temporal-difference residuals with an exponentially decaying weight $\lambda \in [0,1]$. GAE defines the advantage estimate as:
\begin{equation*}
\hat{A}_t^{\mathrm{GAE}(\gamma,\lambda)} = \sum_{l=0}^{\infty} (\gamma \lambda)^l \, \delta_{t+l}^V, \qquad \delta_t^V = r_t + \gamma V(s_{t+1}) - V(s_t),
\end{equation*}
where $\delta_t^V$ is the one-step temporal-difference (TD) error. Lower values of $\lambda$ reduce variance by relying more on shorter-horizon TD estimates, while higher values reduce bias by incorporating longer returns. Proximal Policy Optimisation (PPO) \citep{schulman2017ppo} commonly employs GAE with $\lambda \approx 0.95$ to stabilise training, providing a robust trade-off between bias and variance.

PPO belongs to the family of \emph{actor-critic} methods, where two neural networks are trained jointly:
\begin{itemize}
    \item The \textbf{actor} represents the policy $\pi_\theta(a|s)$, which outputs a distribution over actions given the current state. In execution, this determines how aggressively to trade at each step.
    \item The \textbf{critic} estimates the value function $V_\phi(s)$, parameterised by $\phi$, which predicts the expected return from state $s$.
\end{itemize}
The critic serves as a baseline for advantage estimation, reducing variance in policy-gradient updates. Concretely, the policy gradient is estimated as
\begin{equation*}
\nabla_\theta J(\theta) \approx \mathbb{E}_{t}\left[ \nabla_\theta \log \pi_\theta(a_t|s_t) \, \hat{A}_t \right],
\end{equation*}
where $\hat{A}_t$ is provided by GAE. The critic's role is to supply $V^\pi(s_t)$ in the advantage calculation, thereby reducing variance compared to pure Monte Carlo returns. This actor-critic loop ensures that PPO can adaptively balance exploration of new execution strategies with exploitation of known good policies.

PPO is particularly well-suited for execution problems. Unlike off-policy 
methods, e.g., \edited{deep Q networks (DQN) and soft actor-critic (SAC)}, that can reuse historical data, PPO is on-policy, 
meaning it learns only from trajectories generated by its current policy. 
While this reduces sample efficiency, it provides more stable learning in 
non-stationary environments where market conditions shift. The clipped 
objective (detailed in 
Section~\ref{subsec:ppo_architectures})  prevents destructively large policy updates that could catastrophically 
degrade execution performance. For execution, where a bad policy update could 
mean \edited{significant} losses if deployed, PPO's conservative update mechanism is 
a significant advantage in ensuring stable progress towards an improved policy.

\subsection{Gymnasium for Executing Optimally (GEO)}\label{subsec:geo}

\subsubsection{Data: Minute Bar Data}\label{subsec:minutebardata}

Our back-testing environment is calibrated using minute-bar data from approximately 400 US equities sourced from Mana Tech \citep{manatechllc} for the entirety of the year 2022. Each bar contains bid/ask/mid quotes and displayed depth, trade prices, sided and hidden volumes, among other features. For each symbol and trading day, the dataset provides the variables listed in Table~\ref{tab:data_fields} (in Appendix~\ref{sec:appendix}).

Data cleaning steps include forward-filling missing quotes (at most one bar), excluding minute intervals with no reported trades when constructing returns, and filtering extreme return outliers. Stocks with more than 7\% missing values were removed to avoid instability. Summary statistics of the resulting dataset are shown in Figure~\ref{fig:clean_data} (in Appendix~\ref{sec:appendix}).

\subsubsection{Daily Analytics}

From the \edited{one-minute level} data we construct a set of daily statistics to be used by \edited{GEO} and by the learning models. These include averages of daily trading volume, spreads, order book depth, and trade count, as well as \citet{parkinson1980extreme} high-low volatility estimates\edited{, defined as $$\widehat{\sigma}^{(n)}_d = \sqrt{\frac{1}{4n\ln 2}
\sum_{i=0}^{n-1}\left[\ln(P^H_{d-i}/P^L_{d-i})\right]^2},$$ over 1-, 2-, and 5-day windows, where intraday high-low ranges are used
to} provide robust volatility estimates less sensitive to bid-ask bounce than 
close-to-close returns. Detailed definitions of these features are provided in Appendix~\ref{sec:appendix}, Table~\ref{tab:daily_features_returned}.

\subsubsection{Environment Design and Gymnasium Integration}

\begin{figure}[t]
    \centering
    \includegraphics[width=0.85\linewidth]{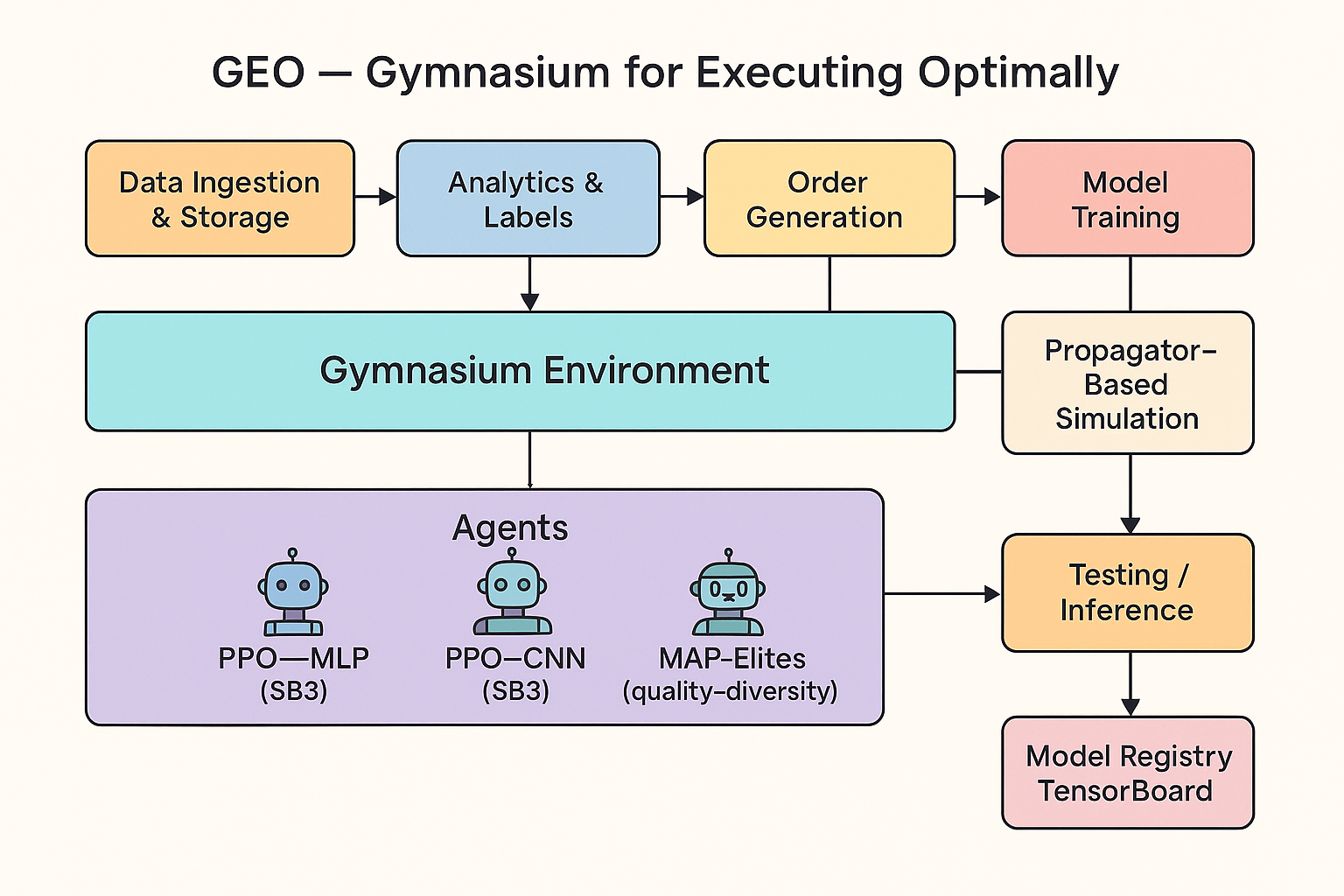}
    \caption{Schematic overview of the GEO environment architecture, showing the 
interaction between agent, environment, and calibrated market impact model.}
    \label{fig:placeholder}
\end{figure}

The \emph{GEO} backtest environment (Figure~\ref{fig:placeholder}) is implemented as a custom Gymnasium environment that adheres to the Gymnasium API \citep{towers2024gymnasium,RLlibGymnasium,SB3Gymnasium}, enabling seamless integration with a wide range of RL algorithms. The design follows three principles:
\edited{
\begin{enumerate}
\item[(i)] \emph{realism}, achieved through historical data-driven calibration of impact and volatility;
\item[(ii)] \emph{modularity}, enabling easy swapping of components such as reward functions or impact models; and
\item[(iii)] \emph{compatibility}, ensuring adherence to Gym's \texttt{reset()} and \texttt{step()} interface.
\end{enumerate}}
Prior researchers have used Gymnasium in simulators such as \emph{ABIDES} and \emph{mbt\_gym}, though their focus has largely been on limit order book (LOB) dynamics \citep{jerome2023mbtgym,amrouni2021abidesgym,hafsi2024optimal}. In contrast, our framework targets execution at the minute-bar horizon with transient market impact.

The core environment extends Gymnasium's \texttt{Env} class and supports vectorised execution for efficient parallel simulation of multiple orders, utilising multiple core CPU acceleration. Each episode corresponds to the execution of a single parent order over $H$ time steps, with each step representing one minute of market time.

The RL execution agent begins from a baseline schedule, defined as a target percentage of expected market volume, and modifies this on a minute-by-minute basis. This design ensures the order is always completed by the horizon $H$. At each step, the agent chooses an action $a_t$ that scales the baseline participation rate up or down. Negative values slow execution, while positive values accelerate it, with $a_t = -1$ corresponding to no trading in that step.

\textbf{Action space:}
\begin{equation*}
a \in \{-1, -0.75, -0.5, -0.25, 0, 0.25, 0.5, 0.75, 1\}
\end{equation*}

This discrete action space simplifies exploration while providing sufficient 
granularity for adaptive scheduling. The symmetric range around zero allows 
both acceleration ($a_t > 0$) and deceleration ($a_t < 0$) from 
the baseline rate, with $a_t = -1$ pausing execution for the current step. 

\textbf{Target rate with action:}
\begin{equation*}
\rho^{\mathrm{action}}_t = \rho^{\mathrm{target}}_t \cdot (1 + a_t), \quad \rho^{\mathrm{target}}_t = \frac{q^{\mathrm{rem}}_t}{\edited{\mathbb{E}}[V_{t,H}]},
\end{equation*}

where $q^{\mathrm{rem}}_t = Q_0 - \sum_{i=0}^{t} q_i$ is the remaining inventory at step $t$, and $\mathbb{E}[V_{t,H}]$ is the expected remaining market volume over the interval $[t,H]$. The executed quantity is then
\begin{equation*}
q_t = \rho^{\mathrm{action}}_t \cdot V^{\mathrm{market}}_t.
\end{equation*}

The environment tracks remaining quantity, elapsed time, arrival slippage, VWAP slippage, holding cost, impact-adjusted fill prices and accumulated impact. State transitions are driven jointly by the agent's actions and exogenous price changes from historical data. 
The agent observes a 13-dimensional feature vector $\mathbf{o}_t \in \mathbb{R}^{13}$ at each timestep, comprising market state, execution progress, impact metrics, and regime context:

\begin{equation*}
\mathbf{o}_t = \begin{bmatrix}
p^{\text{mid}}_t, & V_t, & H-t, & q^{\text{rem}}_t, & \text{ADV\%}, & \text{EHV\%}, \\
p^{\text{fill}}_{t-1}, & q_{t-1}, & I^{\text{imm}}_t, & I^{\text{cum}}_t, \\
p^{\text{arrival}}, & \sigma^{(1)}, & \sigma^{(5)}
\end{bmatrix}^T
\end{equation*}

where $p^{\text{mid}}_t$ is the current mid-price, $V_t$ is market volume at time $t$, $H-t$ is remaining time steps, $q^{\text{rem}}_t$ is remaining inventory, ADV\% and EHV\% are order size relative to average daily volume and expected horizon volume, $p^{\text{fill}}_{t-1}$ and $q_{t-1}$ capture the most recent trade, $I^{\text{imm}}_t$ and $I^{\text{cum}}_t$ are immediate and accumulated impact costs (in basis points), $p^{\text{arrival}}$ is the order arrival benchmark, and $\sigma^{(1)}, \sigma^{(5)}$ are 1-day and 5-day Parkinson volatility estimates.

\subsubsection{Order Generation}\label{subsubsec:order_gen}

Orders are generated using an order generator inside GEO, which samples from historical order flow characteristics to produce a diverse set of parent orders. Each order is parameterised by:
\begin{itemize}
    \item \textbf{Symbol} and \textbf{date}, drawn from the historical dataset.
    \item \textbf{Time horizon}, randomly sampled between 1 and 390 minutes (up to a full trading day).
    \item \textbf{Order size}, expressed as a percentage of expected horizon volume (EHV), sampled from a configurable distribution to simulate varying levels of urgency.
    \item \textbf{Side} ($+1$ for buy, $-1$ for sell), chosen with equal probability.
\end{itemize}

This sampling procedure produces a heterogeneous set of parent orders that vary in size, duration, and liquidity environment, thereby exposing the agent to a broad distribution of execution scenarios. The training dataset is drawn from a historical period strictly preceding the test dataset, ensuring that evaluation is performed under market conditions not encountered during training. This split allows a fair out-of-sample assessment of generalisation performance.

\begin{figure}[p]
  \centering
  \includegraphics[width=\textwidth]{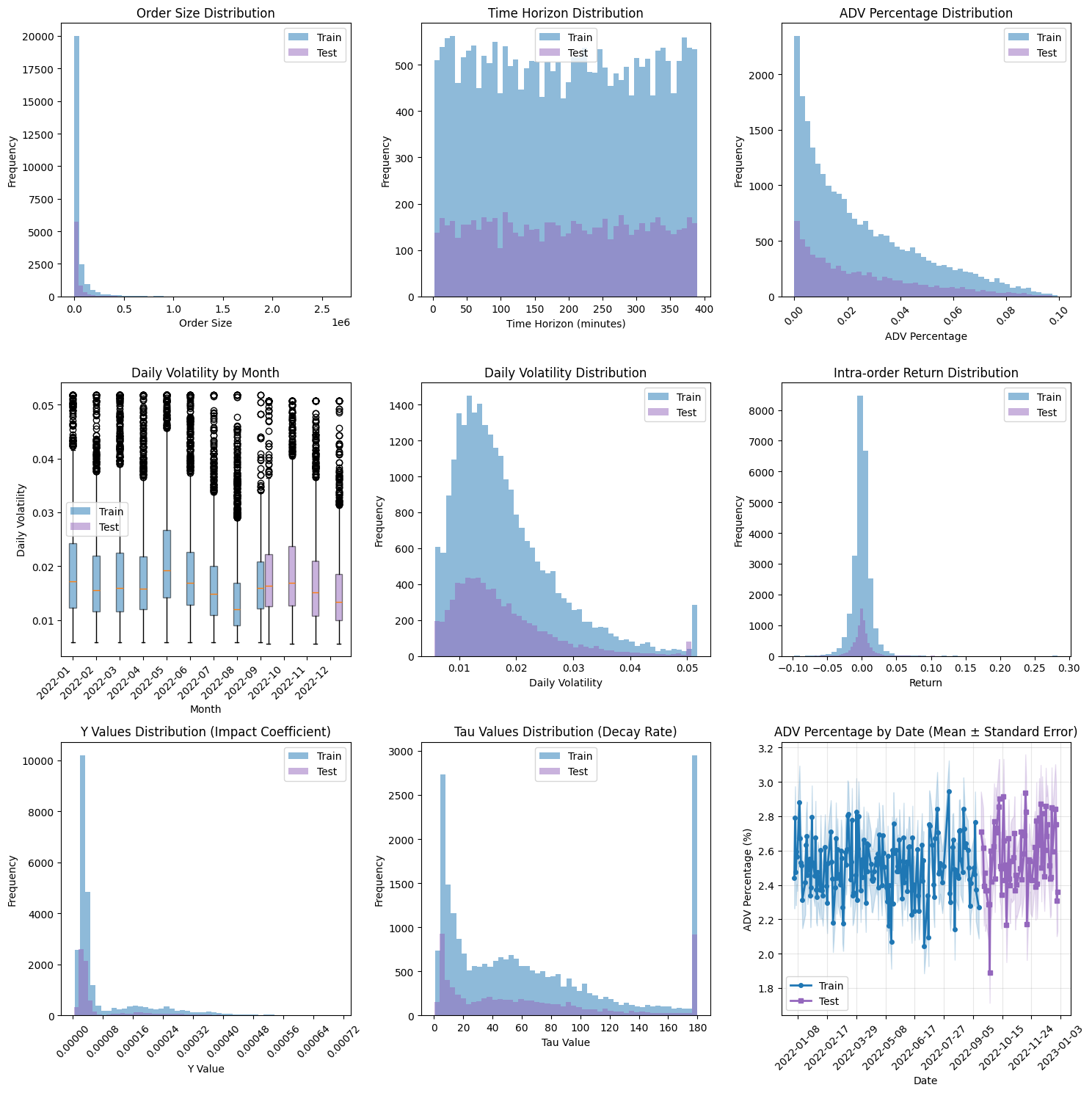}
  \caption{Sample set of generated orders with impact and decay coefficients, stratified by symbol, size, and time horizon.}
  \label{fig:orders_plot}
\end{figure}

In this project, we use nine months of trading data for training and three months for testing, across the universe in the minute bar dataset defined in \edited{S}ection~\ref{subsec:minutebardata}. An illustration of the resulting order size and horizon distributions is provided in Figure~\ref{fig:orders_plot}.

\subsubsection{Execution Simulation}\label{subsec:executionsim}

 At each step $t$, the agent selects an action $a_t$ that scales the 
baseline participation rate to determine trade size $q_t$, which is then 
executed at impact-adjusted prices using the transient impact framework 
(Section~\ref{subsec:propagator}).

The remaining inventory evolves as
\begin{equation*}
q^{\mathrm{rem}}_t = Q_0 - \sum_{i=0}^{t-1} q_i,
\end{equation*}
and the executed quantity in step $t$ is
\begin{equation*}
q_t = \bigl(1 + a_t \bigr)\,\frac{q^{\mathrm{rem}}_t}{\edited{\mathbb{E}}[V_{t,H}]}\,V^{\mathrm{market}}_t,
\end{equation*}
which ensures that the order is fully completed by the horizon $H$.

The execution price is then determined using the propagator model. Specifically, fills are taken at the contemporaneous market VWAP shifted by the cumulative transient impact $I_t$ from current and past trades:
\begin{equation*}
p^{\mathrm{fill}}_t = p^{\mathrm{VWAP}}_t \,\bigl(1 + \mathrm{side} \cdot I_t \bigr),
\end{equation*}
where $\mathrm{side} \in \{+1,-1\}$ indicates buy or sell.

The environment records both the impact-adjusted execution price $p^{\mathrm{fill}}_t$ and the realised implementation shortfall at each step, enabling post-hoc analysis of execution quality and fair comparison across policies. Impact model parameters $(\gamma, G_0, \tau)$ are pre-calibrated 
per stock (Section~\ref{subsec:im_calibration_results}) and assigned to orders 
based on their symbol, ensuring consistent dynamics across training and evaluation.

\subsubsection{Reward Function}\label{subsec:reward}

The agent's reward is the negative of a weighted sum of execution costs:
\begin{equation*}
r_t = -\bigl(\beta_1 C_{\mathrm{arrival}} + \beta_2 C_{\mathrm{VWAP,spread}} + \beta_3 \Delta + \beta_4 \zeta \bigr),
\end{equation*}
where \edited{$\beta_{1},\ldots,\beta_4$} are researcher-defined weights that reflect the relative priority of each objective. In our experiments, $\beta_{1..4}$ are weights reflecting objective priorities. We set 
$\beta_1 = \beta_2 = \beta_3 = 1.0$ and $\beta_4 = 0.1$, emphasizing slippage 
minimization (arrival and VWAP) and schedule adherence while applying a smaller 
weight to the completion penalty for numerical balance.

\begin{tcolorbox}[title=\textbf{Arrival slippage} $C_{\mathrm{arrival}}$, colback=blue!3, colframe=blue!25!black]
\begin{equation*}
C_{\mathrm{arrival}} = \mathrm{side}\,\Bigl(\frac{\sum_{\tau=0}^{t} p^{\mathrm{fill}}_\tau\,|q_\tau|}{\sum_{\tau=0}^{t} |q_\tau|} - p_0\Bigr).
\end{equation*}
Execution shortfall vs. arrival mid-price $p_0$.
\end{tcolorbox}

\begin{tcolorbox}[title=\textbf{VWAP slippage} $C_{\mathrm{VWAP}}$, colback=blue!3, colframe=blue!25!black]
\begin{equation*}
C_{\mathrm{VWAP}} = \mathrm{side}\,\Bigl(\frac{\sum_{\tau=0}^{t} (p^{\mathrm{fill}}_\tau)\,|q_\tau|}{\sum_{\tau=0}^{t} |q_\tau|} - P^{\mathrm{VWAP}}\Bigr).
\end{equation*}
VWAP slippage allows us to measure execution efficiency.
\end{tcolorbox}

\begin{tcolorbox}[title=\textbf{Schedule Deviation} $\Delta$, colback=blue!3, colframe=blue!25!black]
\begin{equation*}
\Delta = \sigma_{\mathrm{minute}}\, \frac{\bigl|\rho_{\mathrm{actual}} - \rho_{\mathrm{target}}\bigr|}{\rho_{\mathrm{target}}}.
\end{equation*}
Volatility-scaled penalty for departing from the target participation rate, where $\sigma_{\mathrm{minute}}$ is the per-minute volatility estimate, computed 
as $\sigma_{\mathrm{daily}}/\sqrt{390}$ using daily Parkinson volatility.
\end{tcolorbox}

\begin{tcolorbox}[title=\textbf{Completion penalty} $\zeta$, colback=blue!3, colframe=blue!25!black]
\begin{equation*}
\zeta = \sigma_{\mathrm{minute}}\, \frac{q_{\mathrm{rem}}}{Q_0}.
\end{equation*}
Discourages unfinished inventory; proportional to $q_{\mathrm{rem}}$ and market turbulence.
\end{tcolorbox}

A negative slippage (buying below the benchmark for buys, or selling above it for sells) contributes positively to the reward via the sign in $C_{\mathrm{arrival}}$ and $C_{\mathrm{VWAP,spread}}$.

\subsubsection{Baseline Strategies}

We construct a set of standard baseline strategies to benchmark the experimental agents against. The implementations for TWAP, VWAP, and POV emulate common execution algorithms, providing realistic reference points. A purely random policy is also included as a noise-driven comparator.

\begin{tcolorbox}[title=\textbf{TWAP -- Time-Weighted}, colback=blue!3, colframe=blue!25!black]
\begin{equation*}
q_t^{\mathrm{TWAP}}=\frac{Q_0}{H}.
\end{equation*}
Constant shares each minute.
\end{tcolorbox}

\begin{tcolorbox}[title=\textbf{VWAP -- Volume-weighted}, colback=blue!3, colframe=blue!25!black]
\begin{equation*}
q_t^{\mathrm{VWAP}} = Q_0 \cdot \frac{\bar V_t}{\sum_{\tau=0}^{H-1}\bar V_\tau}.
\end{equation*}
Allocate proportionally to historical intraday volume profile $\bar V_t$, 
where $\bar V_t$ is the average market volume at minute $t$ over the past N 
trading days.
\end{tcolorbox}

\begin{tcolorbox}[title=\textbf{POV -- Percentage of Volume}, colback=blue!3, colframe=blue!25!black]
\begin{equation*}
q_t^{\mathrm{POV}} = \rho_{\mathrm{target}}\, V_t, \qquad \rho_{\mathrm{target}} = \frac{Q_0}{\sum_{\tau=0}^{H-1}\bar V_\tau}.
\end{equation*}
Trade a fixed participation of \emph{realised} market volume $V_t$; $\rho_{\mathrm{target}}$ chosen to complete in expectation over $H$ based on historical volume $\bar V[a,H]$.
\end{tcolorbox}

\begin{tcolorbox}[title=\textbf{Random -- Noise baseline}, colback=blue!3, colframe=blue!25!black]
At each minute draw $a_t$ from the action set and set
\begin{equation*}
q_t^{\mathrm{RAND}} = (1+a_t)\,\frac{q^{\mathrm{rem}}_t}{\mathbb{E}[V_{t,H}]}\,V_t, \quad q^{\mathrm{rem}}_t = Q_0 - \sum_{i=0}^{t-1} q_i,
\end{equation*}
so completion is still enforced by the target-rate scaffolding.
\end{tcolorbox}

All baselines are evaluated within the GEO environment using the same calibrated 
transient impact model and fill price mechanism as the RL agents, ensuring fair 
comparison. Each baseline executes orders according to its deterministic schedule, 
with fills adjusted for market impact via Equation~\eqref{eq:propagator_full}.

\subsection{Novel RL Approaches to Optimal Execution}

\subsubsection{Proximal Policy Optimisation models (PPO)}\label{subsec:ppo_architectures}

PPO belongs to the family of \emph{policy-gradient methods}, which directly optimise a parametrised policy $\pi_\theta(a|s)$ by estimating gradients of the expected return with respect to the parameters $\theta$. Unlike value-based methods such as Q-learning, which learn an action-value function $Q(s,a)$ and act greedily via $\arg\max_a Q(s,a)$, policy-gradient methods adjust $\pi_\theta$ itself to increase the probability of selecting actions that yield higher returns. Formally,
\begin{equation*}
\nabla_\theta J(\pi_\theta) = \mathbb{E}_{s_t \sim d^{\pi_\theta}, a_t \sim \pi_\theta}\left[\nabla_\theta \log \pi_\theta(a_t|s_t)\, Q^{\pi_\theta}(s_t,a_t)\right].
\end{equation*}

PPO is an \emph{on-policy} algorithm, meaning it learns exclusively from trajectories generated by its current policy. This stabilises training but requires discarding past data once the policy updates. By contrast, off-policy methods such as Q-learning or SAC update from a behaviour policy $\mu \neq \pi_\theta$, reusing past trajectories to improve sample efficiency.

The key innovation of PPO is its \emph{clipped surrogate loss}, which stabilises learning by preventing excessively large policy updates. With generalised advantage estimation (GAE), the PPO objective is
\begin{equation*}
L^{\text{CLIP}}(\theta) = \mathbb{E}_t \left[ \min \Big( r_t(\theta) \, \hat{A}_t, \operatorname{clip}\big(r_t(\theta), 1-\epsilon, 1+\epsilon\big)\,\hat{A}_t \Big) \right],
\end{equation*}
where
\begin{equation*}
r_t(\theta) = \frac{\pi_\theta(a_t|s_t)}{\pi_{\theta_{\text{old}}}(a_t|s_t)}.
\end{equation*}

Here, $\hat{A}_t$ denotes the estimated advantage of action $a_t$ at state $s_t$, and $\epsilon$ (typically $0.1$-$0.3$) controls the clipping range. The clipping prevents $r_t(\theta)$ from deviating too far from $1$, thereby limiting the size of policy updates. PPO also incorporates a separate value-function loss and an entropy bonus to balance exploitation and exploration.

In addition to the clipped policy objective, PPO optimises a state-value baseline $V_\phi(s_t)$ using a squared-error loss:
\begin{equation*}
L^{\mathrm{VF}}(\phi) = \mathbb{E}_t \Big[ \big(V_\phi(s_t) - \hat{R}_t\big)^2 \Big],
\end{equation*}
where $\hat{R}_t$ is the empirical return or bootstrapped target. This helps reduce variance in policy-gradient updates by ensuring the critic approximates the long-term execution cost of continuing from state $s_t$.

To avoid premature convergence to deterministic (and possibly suboptimal) policies, PPO adds an entropy regularisation term:
\begin{equation*}
L^{\mathrm{entropy}}(\theta) = \mathbb{E}_t \big[ \mathcal{H}(\pi_\theta(\cdot|s_t)) \big],
\end{equation*}
where $\mathcal{H}$ is the Shannon entropy. This encourages exploration by rewarding policies that maintain uncertainty over possible actions.

Putting it all together, the PPO loss combines the clipped surrogate policy objective with a value-function regression term and an entropy bonus:
\begin{equation}
\label{eq:ppo_master}
\begin{aligned}
\mathcal{L}_{\mathrm{PPO}}(\theta,\phi) = \mathbb{E}_t\Big[ & \underbrace{\min\!\Big(r_t(\theta)\,\hat{A}_t, \mathrm{clip}\!\big(r_t(\theta),1-\epsilon,1+\epsilon\big)\,\hat{A}_t\Big)}_{L^{\pi}} \\
& - c_v \,\underbrace{\big(V_\phi(s_t)-\hat{R}_t\big)^2}_{L^V} + c_e \,\underbrace{\mathcal{H}\!\big(\pi_\theta(\cdot\mid s_t)\big)}_{L^H} \\
& - c_{\mathrm{KL}}\,\underbrace{\mathrm{KL}\!\Big(\pi_{\theta_{\mathrm{old}}}(\cdot\mid s_t)\|\pi_\theta(\cdot\mid s_t)\Big)}_{L^{KL}} \Big],
\end{aligned}
\end{equation}

We implement PPO agents with two distinct feature extractor architectures. The first is an multilayer perceptron (MLP) extractor, which normalises the flattened observation vector and passes it through two fully connected layers with ReLU activations. This design is lightweight and fast, making it well-suited to compact, tabular state representations.  The second applies one-dimensional 
convolutions across the feature dimension with three blocks (64, 64, 128 channels), 
SiLU activations, and group normalization, providing additional representational 
capacity at the cost of increased computation.

\textbf{Observation Preprocessing:} Raw observations are normalised using 
running mean and standard deviation estimates maintained across vectorised 
environments (Stable-Baselines3 VecNormalize wrapper), ensuring numerical 
stability across assets and regimes.

Hyperparameters follow standard PPO best practices \citep{schulman2017ppo} with 
minor adjustments for the execution domain.

\begin{tcolorbox}[title=\textbf{Common PPO Settings}, colback=blue!3, colframe=blue!25!black]
\textbf{Architecture:} Shared feature extractor (256-dim); actor MLP [256, 256, 128]; critic MLP [256, 256, 128]; tanh activation.\\[4pt]
\textbf{Optimiser:} PPO with clipped surrogate loss $L^\pi$, value loss $L^V$, entropy bonus $L^H$, and KL penalty $L^{KL}$ (Eq.~\ref{eq:ppo_master}).\\[4pt]
\textbf{Key Hyperparameters:}
\begin{itemize}
    \item Rollout: $n_{\mathrm{steps}}=2048$ per environment, discount $\gamma=0.999$
    \item Batch size: Auto-scaled by rollout size $\in\{2048, 4096, 8192\}$
    \item Training: 3 epochs, clip range $\epsilon \approx 0.18$ (linear decay)
    \item Regularization: target KL = 0.02, entropy coefficient = 0.006
    \item Value loss coefficient = 0.55, max gradient norm = 0.5
\end{itemize}
\textbf{Learning Rate:} Linear decay from $3 \times 10^{-4}$ to 0.\\[4pt]
\textbf{GAE:} $\lambda = 0.95$ for advantage estimation.
\end{tcolorbox}

\subsubsection{MAP-Elites: Quality-Diversity Optimisation}\label{subsec:mapelites}

Most reinforcement learning algorithms optimise for a single performance objective, aiming to find the policy $\pi_\theta$ that maximises expected return. While this produces a single ``best'' solution under the training reward, it provides little insight into the diversity of alternative strategies or their robustness under changing market conditions. The \emph{MAP-Elites} algorithm, introduced by \citet{mouret2015mapelites}, belongs to the class of \emph{quality-diversity} (QD) methods. Rather than converging on a single optimum, MAP-Elites searches for a collection of high-performing yet behaviourally distinct solutions, yielding a repertoire of policies that together ``illuminate'' the range of possible strategies.

\begin{tcolorbox}[title=\textbf{MAP-Elites Algorithm (overview)}, colback=blue!3, colframe=blue!25!black]
\textbf{Inputs:} behaviour-descriptor mapping $b(\pi)\in\mathbb{R}^d$, archive grid $\edited{\mathcal{G}}$ partitioning descriptor space, quality function $Q(\pi)$, variation operator $\mathrm{Var}(\cdot)$.

\begin{enumerate}
  \item \textbf{Initialisation.} Generate candidate policies $\{\pi^{(i)}\}$. For each, evaluate $Q(\pi^{(i)})$ and compute descriptor $z^{(i)}=b(\pi^{(i)})$.
  \item \textbf{Archive update.} Map $z^{(i)}$ to grid cell $c$. If $\edited{\mathcal{G}}[c]$ is empty, insert $\pi^{(i)}$; if occupied, replace only if $Q(\pi^{(i)})$ exceeds the incumbent via the quality function.
  \item \textbf{Variation.} Sample elites from $\edited{\mathcal{G}}$ and generate offspring by perturbing their representation.
  \item \textbf{Iteration.} Evaluate offspring, compute descriptors, and update $\edited{\mathcal{G}}$. Iterate until evaluation budget is exhausted.
\end{enumerate}

\textbf{Output:} repertoire $\edited{\mathcal{G}}$ of elites $\{\pi^\star_c\}$, one per descriptor cell, each maximising $Q(\pi)$ locally.
\end{tcolorbox}

The quality function $Q(\pi)$ is the negative mean total execution cost averaged 
over evaluation episodes, where lower cost (better execution) yields higher quality. 
Each policy is evaluated on orders matching its phenotype to ensure fair niche-specific 
comparison.

When applying variation to neural networks, we perturb the weights $\theta$ that define the extractors allowing for variability. Offspring are generated by applying Gaussian 
noise ($\mathcal{N}(0, \sigma^2)$) to parent policy parameters. This preserves learned behaviors while introducing 
local variation.

Each cell in the archive corresponds to a region of descriptor space. MAP-Elites guarantees that only the best-known policy for each behavioural niche is retained. Over many iterations, the archive becomes populated with a structured set of strategies that are both high quality and diverse.

In our implementation, descriptors are designed to capture the fundamental market dimensions of \emph{liquidity} and \emph{volatility}. Both are normalised by their empirical quantiles across the universe, mapping values to the unit interval $[0,1]$ in a rank-preserving way:

\begin{itemize}
  \item \textbf{Liquidity.} Quantile-normalised average daily volume (ADV) of symbol $S$:
  \[
  b^{\mathrm{liq}}(S) = \frac{1}{N}\sum_{j=1}^N \mathbf{1}\{\mathrm{ADV}(S_j) \leq \mathrm{ADV}(S)\},
  \]
  where $N$ is the number of symbols in the universe.

  \item \textbf{Volatility.} Quantile-normalised one-day Parkinson estimator:
  \[
  b^{\mathrm{vol}}(S) = \frac{1}{N}\sum_{j=1}^N \mathbf{1}\{\widehat{\sigma}^{(1)}(S_j) \leq \widehat{\sigma}^{(1)}(S)\},
  \]
  where $\widehat{\sigma}^{(1)}(S) = \sqrt{\tfrac{1}{4\ln 2}\left[\ln\!\left(\tfrac{P^H}{P^L}\right)\right]^2}$.
\end{itemize}

Thus, each policy $\pi_\theta$ is mapped into the two-dimensional descriptor space
\[
b(\pi_\theta) = \big(b^{\mathrm{liq}}(S), b^{\mathrm{vol}}(S)\big) \in [0,1]^2,
\]
with archive cells representing liquidity-volatility niches.

\begin{figure}[t]
\centering
\begin{tikzpicture}[x=0.8cm,y=0.8cm,>=Latex]
  \def\N{6}
  \node[anchor=east] at (-0.6, \N/2) {\small Volatility};
  \node at (\N/2,-0.8) {\small Liquidity};
  \draw[->,thick] (0,0) -- (\N+0.6,0);
  \draw[->,thick] (0,0) -- (0,\N+0.6);
  \foreach \i in {0,...,\N} {
    \draw[gray!40] (\i,0) -- (\i,\N);
    \draw[gray!40] (0,\i) -- (\N,\i);
  }
  \foreach \x/\y/\col/\q in {
    1/1/green!35/{Q=0.58},
    2/4/green!25/{Q=0.53},
    3/2/green!45/{Q=0.61},
    4/5/green!55/{Q=0.67},
    5/1/green!30/{Q=0.50},
    5/3/green!60/{Q=0.70}
  }{
    \filldraw[fill=\col,draw=black!40] (\x,\y) rectangle ++(1,1);
    \node[font=\scriptsize] at (\x+0.5,\y+0.5) {\q};
  }
  \draw[black!40,fill=green!55] (\N+1, \N-0.2) rectangle ++(0.5,0.5);
  \node[anchor=west, font=\scriptsize] at (\N+1.6, \N+0.05) {Elite};
  \draw[black!40] (\N+1, \N-1.0) rectangle ++(0.5,0.5);
  \node[anchor=west, font=\scriptsize] at (\N+1.6, \N-0.75) {Empty cell};
\end{tikzpicture}
\caption{MAP-Elites archive across liquidity and volatility descriptors. Each filled cell stores the highest-quality policy (elite) discovered in that region.}
\label{fig:map_elites_grid}
\end{figure}
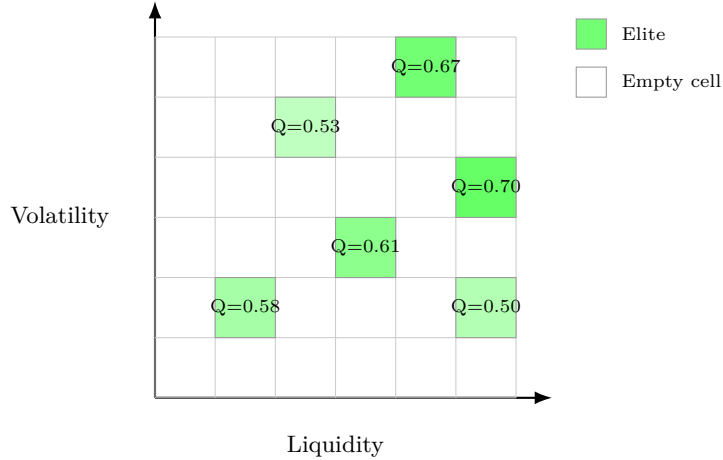

MAP-Elites offers potential advantages: it may enforce exploration of behaviourally 
distinct strategies, produce an interpretable map of ``what works where,'' provide 
robustness through diverse repertoires, and remain agnostic to policy representation. 
In this work, we apply MAP-Elites to a PPO-based CNN policy using liquidity-volatility 
descriptors to structure the archive. However, as we show in Section~\ref{sec:results}, 
the practical benefits depend critically on descriptor choice, fitness function 
design, and computational budget.

\section{Results}\label{sec:results}

\subsection{Transient Impact Model Calibration}\label{subsec:im_calibration_results}

We calibrate the propagator model parameters $(G_0,\tau)$ by regressing returns $r_t$ onto lagged signed volumes $\epsilon_{t-\ell} f(v_{t-\ell})$ across multiple lags $\ell=1,\dots,L$. Parameters are chosen to maximise out-of-sample $R^2$ across rolling windows of \edited{one-minute level} data for 400+ S\&P 500 stocks during 2022 (excluding stocks with insufficient or missing 
\edited{one-minute level} data).

Two functional forms for $f(v)$ were compared: linear and square-root. Consistent with empirical microstructure studies, the square-root form achieved higher mean $R^2$ across symbols and was adopted as the baseline:
\begin{equation*}
f(v_t, q_t) = \gamma \sqrt{\tfrac{|q_t|}{v_t}},
\end{equation*}

where $q_t$ is the signed traded quantity at time $t$, and $v_t$ is the market volume for the period $t$ and $\gamma$ is the impact coefficient. 

The decay kernel $G(\ell)$ captures temporal persistence of impact:
\begin{equation*}
G(\ell) = G_0 \, e^{-\ell / \tau},
\end{equation*}
where $G_0$ is the immediate impact coefficient, $\ell$ is the lag since the trade, and $\tau$ the characteristic decay horizon.

\begin{figure}[t]
    \centering
    \includegraphics[width=0.9\linewidth]{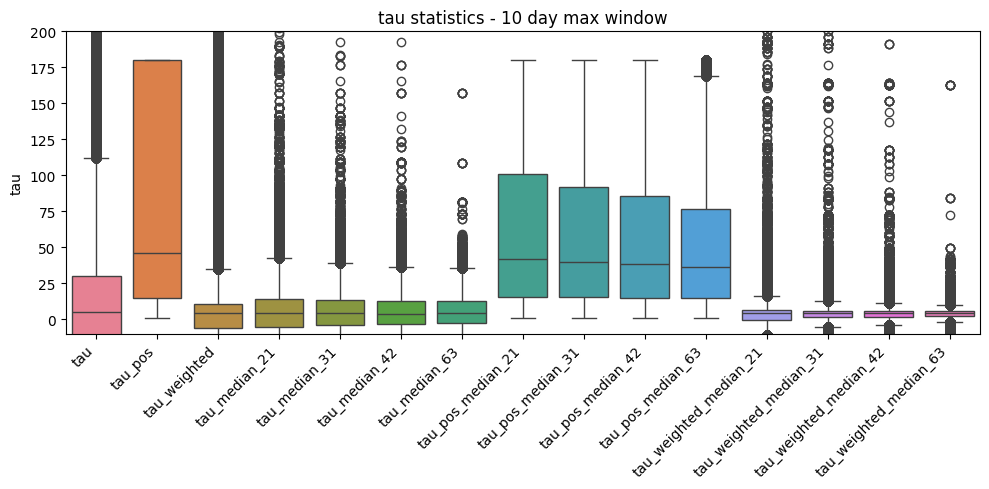}
    \caption{Constraint necessity: without bounds, $\tau$ can turn negative due to noise (Eq.~\ref{eq:propagator_full}), implying explosion rather than decay.}
    \label{fig:tau_samples}
\end{figure}

Calibration is performed via constrained non-linear least squares with economically interpretable bounds:
\begin{equation*}
G_0 \geq 0, \qquad \tau \in [0.5, 180].
\end{equation*}
The necessity of these contraints is demonstrated in Figure~\ref{fig:tau_samples}.

\begin{figure}[t]
  \centering
  \includegraphics[width=\linewidth]{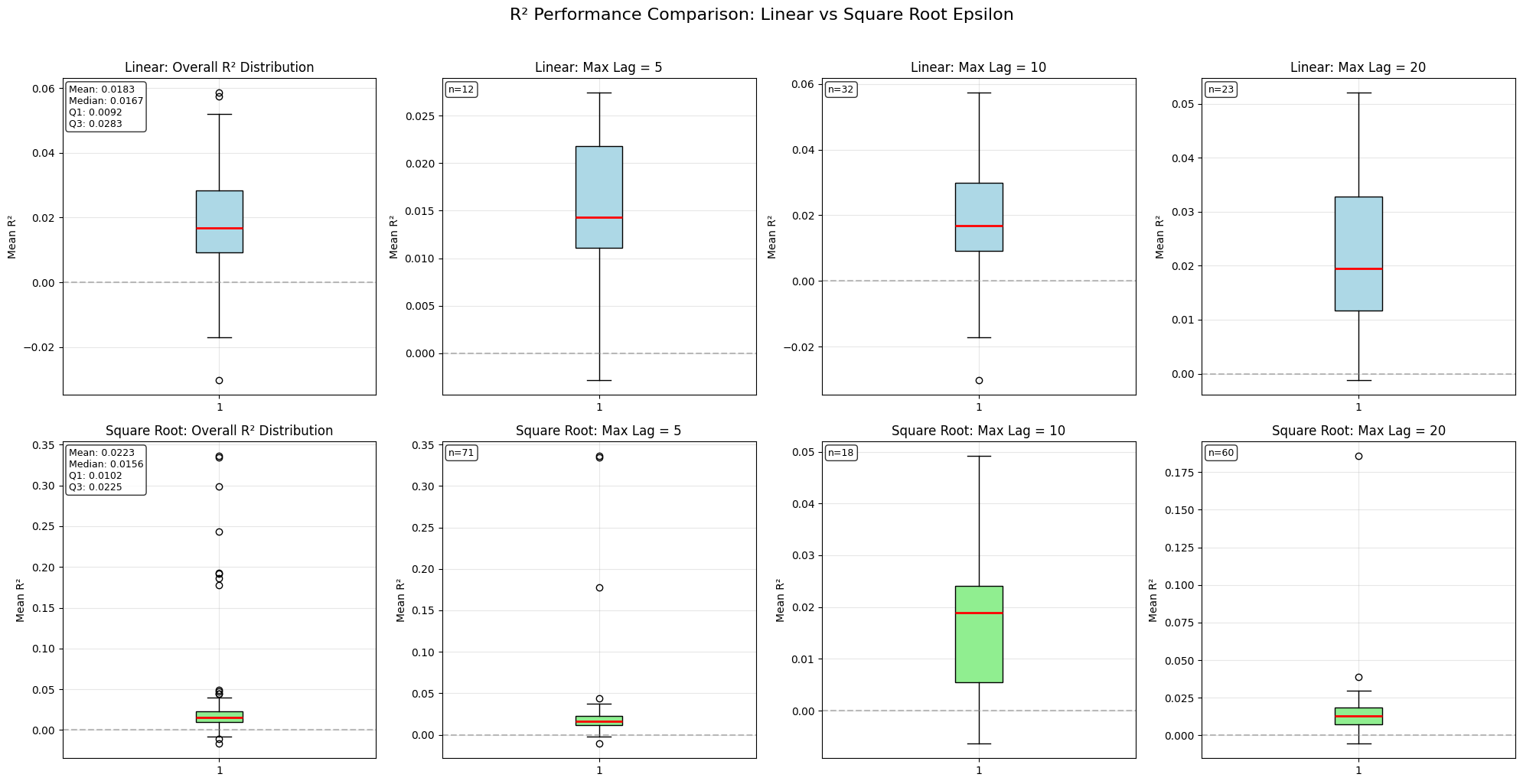}
  \caption{$R^2$ comparison of linear vs. square-root instantaneous impact across lags $\ell=1,\dots,L$ for $L \in \{5, 10, 20, 30\}$. Square-root dominates in liquid names and shorter horizons.}
  \label{fig:lag_study}
\end{figure}

\begin{figure}[t]
  \centering
  \includegraphics[width=0.7\textwidth]{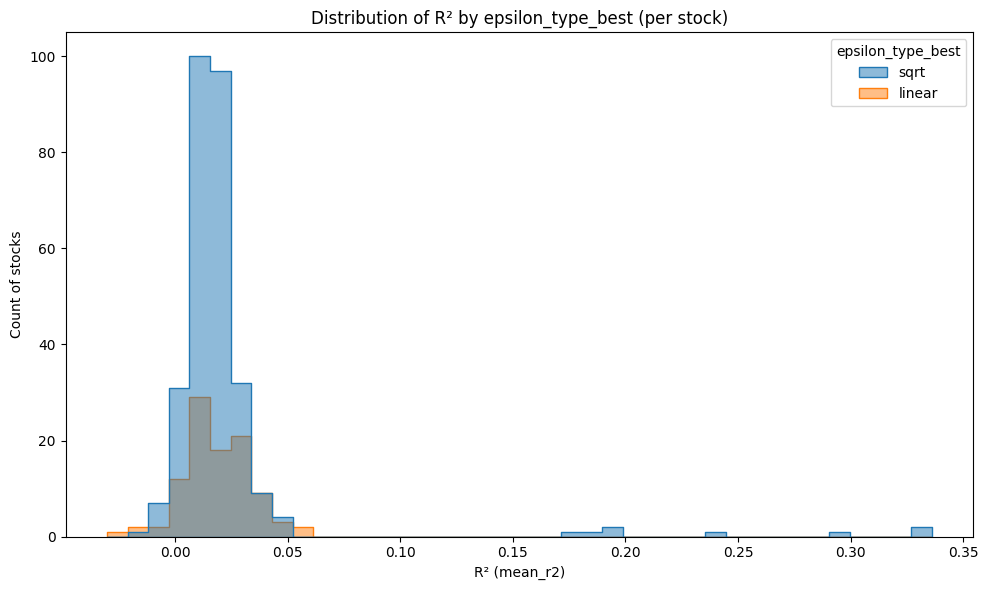}
  \caption{Aggregate $R^2$ outcomes: the square-root specification outperforms linear on average, supporting its choice as the baseline model.}
  \label{fig:sqrt_vs_linear}
\end{figure}

Figure~\ref{fig:lag_study} presents a lagged regression study comparing $R^2$ of linear and square-root forms at maximum lags $L$ of 5, 10, and 20 minutes. The square-root function dominates overall, especially for liquid names where concavity captures diminishing marginal impact (Figure~\ref{fig:sqrt_vs_linear}).

After cross-validation, we store the best $(\ell,\tau,\gamma)$ and \edited{average} $\bar{R}^2$ for reuse, retaining only symbols with $\bar{R}^2 > 0.02$. This threshold leaves about two-thirds of the universe available for training and testing. Stocks below this threshold are excluded from the training, validation and test sets.

\subsection{RL Performance Results}

\subsubsection{Evaluation of RL models}

For evaluation, we adopt a strict train-test separation based on non-overlapping calendar periods. The training set consists of orders generated between January 1, 2022 and September 30, 2022, while the evaluation set spans October 1 to December 31, 2022.

Training is performed within the GEO environment, which integrates Gymnasium for vectorised simulation. For each experiment, we generate a large set of synthetic parent orders (5,000-25,000 per run) sampled across the universe described. Orders are balanced between buys and sells and cover a range of sizes and horizons. The exact same training and evaluation orders are shared across all policies to guarantee a fair comparison.

We benchmark PPO with multi-layer perceptron (MLP) and convolutional (CNN) feature extractors, MAP-Elites, and baseline strategies (TWAP, VWAP, POV, and Random). To mitigate the influence of outliers, we exclude pathological orders and winsorise evaluation metrics at the 1st and 99th percentiles before aggregation.

\subsubsection{PPO Results and Findings}

Table~\ref{tab:ppo_exec_summary} summarises performance for the two PPO architectures compared to baseline strategies. Metrics are expressed in basis points (bps), with standard errors in parentheses.

\begin{table*}[t]
\centering
\resizebox{\textwidth}{!}{
\begin{tabular}{lccccccc}
\toprule
Agent & Count & Notional (Bln) & Arrival Slippage & Duration \% & Return & Total Cost & Action \% \\
\midrule
ppo\_mlp & 4900 & 21.32 & 3.78 (0.93) & 99.2 & -5.19 (1.41) & 178.26 (1.77) & 18.25 (0.06) \\
vwap & 4900 & 21.66 & 5.23 (1.01) & 99.2 & -5.60 (1.43) & 476.11 (6.09) & 12.51 (0.08) \\
random & 4900 & 21.31 & 3.77 (0.96) & 99.2 & -3.46 (1.41) & 217.58 (2.23) & -0.02 (0.06) \\
pov & 4900 & 21.30 & 4.07 (0.97) & 99.3 & -3.87 (1.41) & 211.71 (2.21) & 0.00 (0.00) \\
twap & 4900 & 20.18 & 7.01 (0.90) & 98.8 & 1.70 (1.40) & 302.89 (3.25) & 75.59 (0.05) \\
ppo\_cnn & 4900 & 21.41 & 2.13 (0.92) & 99.2 & -5.68 (1.41) & 178.70 (1.78) & 19.00 (0.06) \\
\bottomrule
\end{tabular}}
\caption{Execution summary for PPO and baseline strategies. Arrival slippage, Return, and Total Cost in basis points (bps); Standard errors in parentheses; Duration \% is the average portion of  of $H$ before completion; Action is the mean $a_t$ over all parents; Return: intra-order price drift; }
\label{tab:ppo_exec_summary}
\end{table*}

The PPO-CNN model achieves the lowest arrival slippage of all strategies (2.13 bps), a statistically significant improvement  where $p<0.05$,  relative to VWAP, TWAP, and POV. PPO-MLP delivers performance comparable to the random baseline on slippage but still far outperforms benchmarks on cost. Both PPO agents reduce total cost dramatically, halving costs relative to TWAP (303 bps) and cutting more than 60\% compared to VWAP (476 bps).

\begin{figure}[t]
    \centering
    \includegraphics[width=\linewidth]{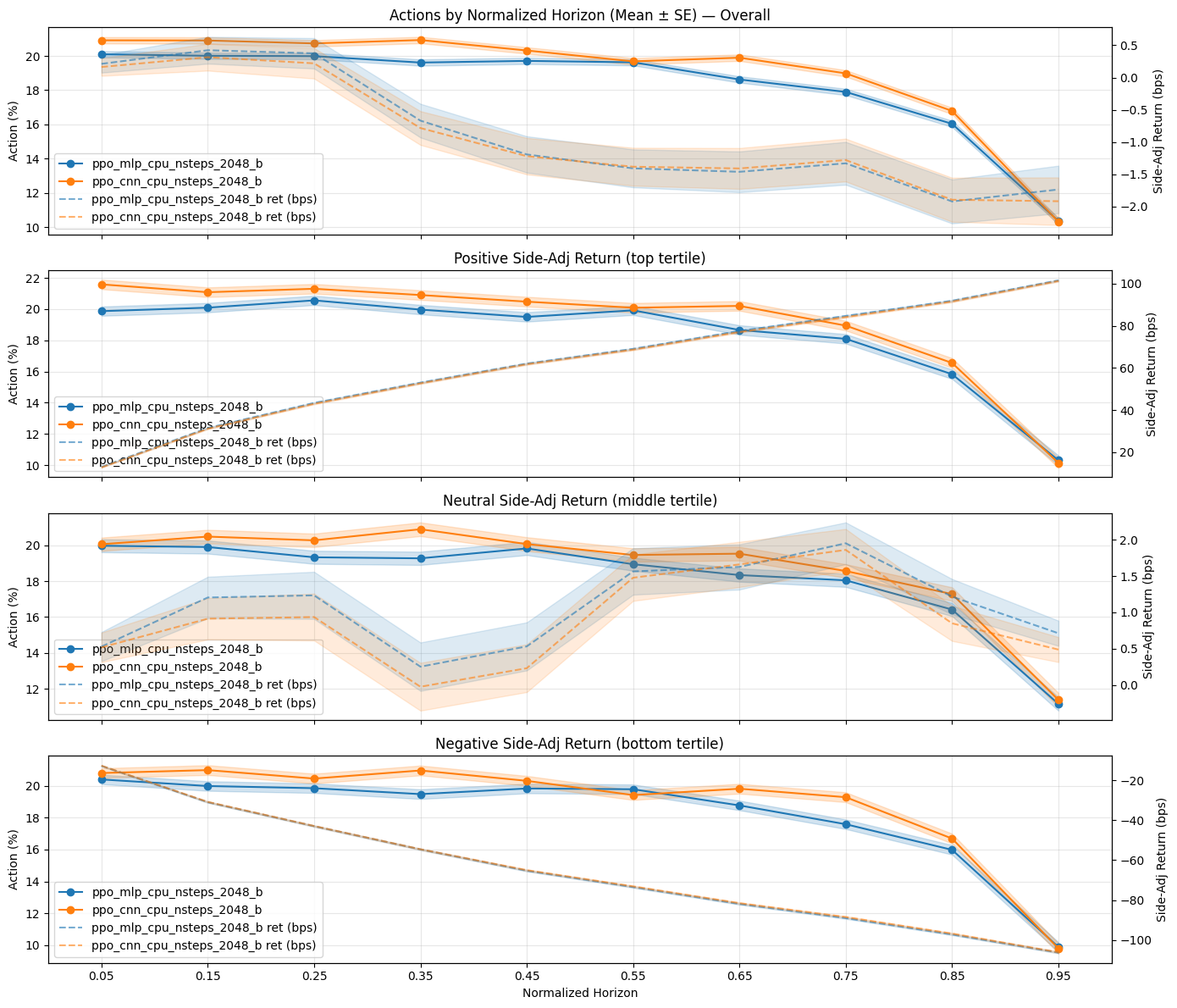}
    \caption{Aggregated mean action \% $a_t$ across order horizons, conditioned on price drift relative to order side. PPO agents exhibit front-loading consistent with Almgren-Chriss scheduling.}
    \label{fig:ppo_trajectories}
\end{figure}

\begin{figure}[t]
    \centering
    \includegraphics[width=0.8\linewidth]{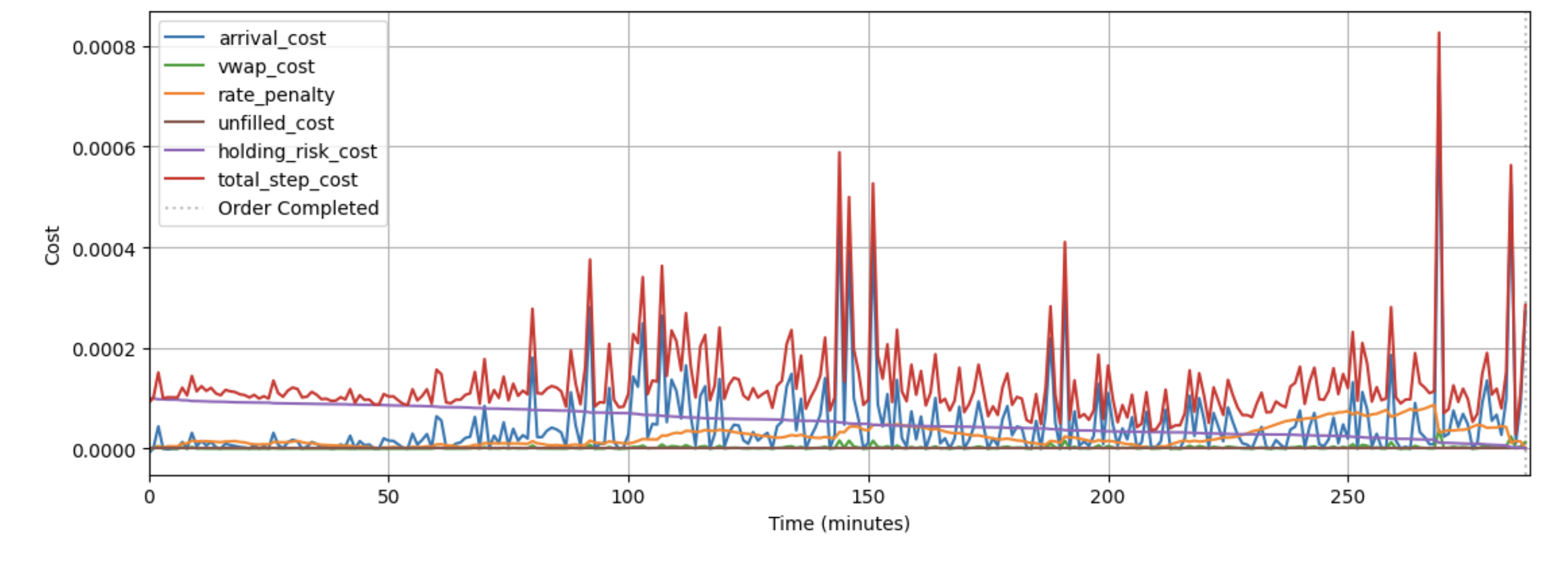}
    \caption{Decomposition of costs across strategies. PPO agents internalise holding cost, which drives front-loading behaviour.}
    \label{fig:rl_costs}
\end{figure}

Both PPO agents adopt a front-loaded pattern consistent with Almgren-Chriss intuition, mitigating holding costs by executing earlier. The CNN variant moderates this front-loading when returns are adverse, suggesting that the temporal structure enables a more nuanced response to price drift (Figures~\ref{fig:ppo_trajectories} and \ref{fig:rl_costs}).

\subsection{MAP-Elites Results and Findings}\label{subsec:mapelites_results}

\subsubsection{Exploratory Application of MAP-Elites}

We conducted a preliminary investigation of MAP-Elites for generating regime-specialist 
policies. While quality-diversity approaches have shown promise in robotics 
\citep{mouret2015mapelites}, their application to financial execution remains 
largely unexplored. We implemented MAP-Elites over volatility (normalised Parkinson) and liquidity (normalised ADV) behavioural descriptors 
with a $3 \times 3$ grid, seeding policies from a baseline PPO-CNN model and 
evolving them via Gaussian parameter perturbations ($\sigma = 0.01$). Initial 
experiments with modest configurations (30-100 iterations) showed promising 
in-sample improvements but failed to generalise. We therefore scaled to 500 
iterations with 256 children per generation, evaluating 128,000 candidate policies 
over 5.5 hours on Apple M4 Max 64GB 16-core CPU.

Table~\ref{tab:mapelites_cells} presents phenotype-specific performance: each 
specialist was evaluated exclusively on test orders matching its liquidity-volatility 
cell. 
Three cells achieved 8-10\% improvements over baseline PPO within their 
training niches, while others showed degradation, particularly in 
low-liquidity regimes. These findings suggest potential for quality-diversity 
methods but indicate that effective deployment requires careful consideration 
of regime boundaries, training data density per cell, and selective 
application strategies.

\begin{figure}[!p]
   \centering
   \includegraphics[width=\linewidth]{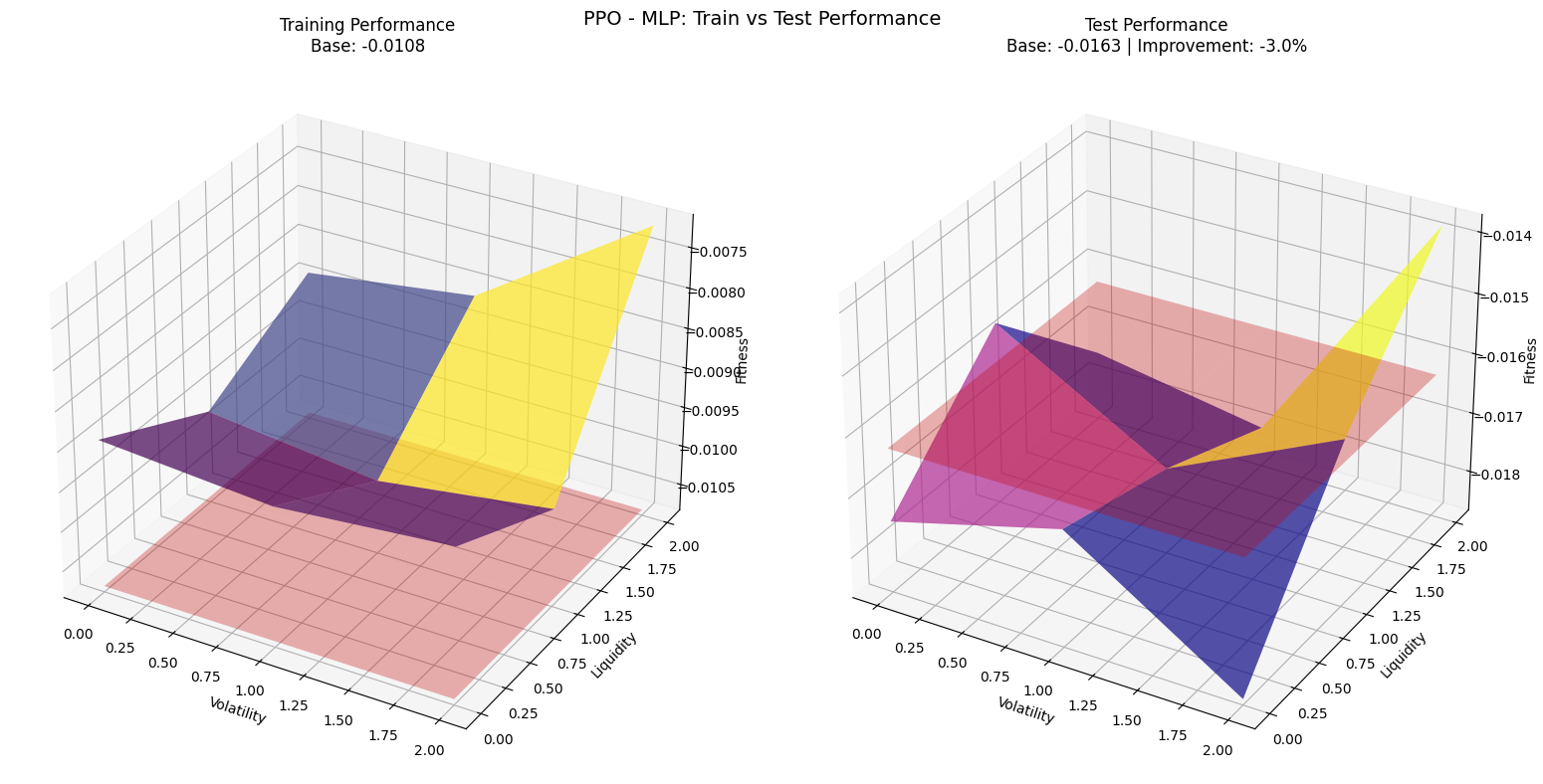}
   \caption{MAP-Elites archive evolution over 100 iterations. Z-axis indicates fitness (negative total cost). Red plane shows baseline PPO-CNN fitness. Training set (left) achieves improvements across all cells, whereas test set (right) shows mixed generalization with failures in low-liquidity regimes. }
   \label{fig:mapelites_3d}
\end{figure}

\begin{figure}[!p]
   \centering
   \includegraphics[width=\linewidth]{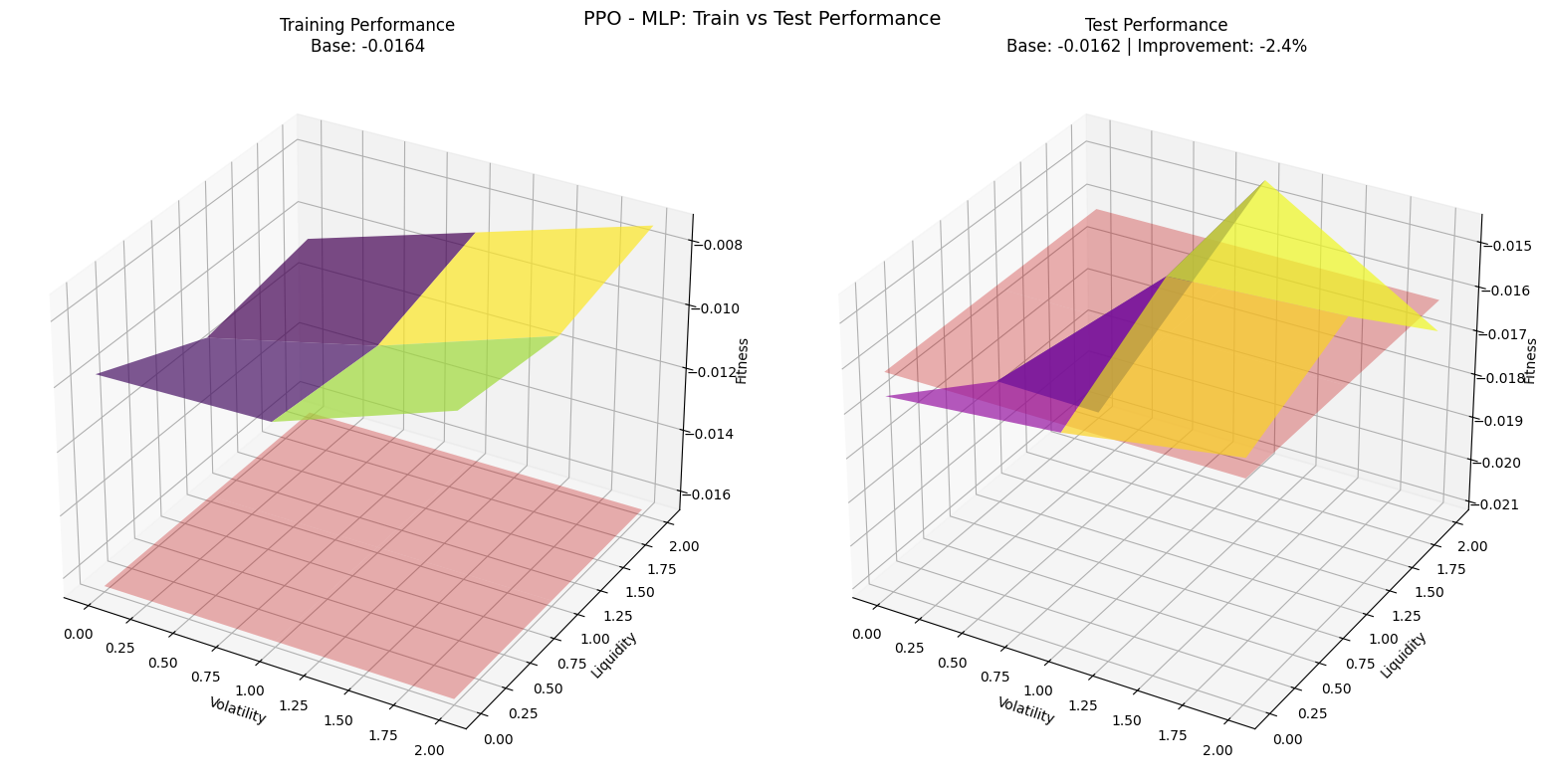}
   \caption{MAP-Elites archive evolution. 500 iteration run. Z-axis indicates fitness (negative total cost). Red plane shows baseline PPO-CNN fitness. Training set (left) achieves improvements across all cells, whereas test set (right) shows improved results vs. the 100 iteration run. }
   \label{fig:mapelites_3d_yolo}
\end{figure}
The archive reveals striking heterogeneity in generalisation. Three cells achieved 8-10\% improvements over baseline, with the high-volatility/medium-liquidity cell performing best at +10.3\%. Conversely, the high-volatility/low-liquidity cell degraded catastrophically (-30.2\%), suggesting overfitting in illiquid regimes. While the overall cell average showed -2.4\% degradation, individual specialists demonstrate that regime-specific policies can outperform universal approaches when properly matched to market conditions. 

These results motivate development of ensemble routing strategies that selectively 
deploy specialists only in regimes where they demonstrate robust out-of-sample 
improvements. Such meta-policies are left for future work.

\begin{table}[t]
\centering
\caption{MAP-Elites specialist fitness (total cost) performance by market regime}
\label{tab:mapelites_cells}
\begin{tabular}{llrr}
\toprule
Volatility & Liquidity & Fitness & vs CNN Policy \\
\midrule
Low & Low & -0.01672 & -3.3\% \\
Low & Medium & -0.01637 & -1.2\% \\
Low & High & -0.01574 & +2.7\% \\
Medium & Low & -0.01834 & -13.3\% \\
Medium & Medium & -0.01487 & +8.1\% \\
Medium & High & -0.01467 & +9.3\% \\
High & Low & -0.02107 & -30.2\% \\
High & Medium & -0.01451 & \textbf{+10.3\%} \\
High & High & -0.01689 & -4.4\% \\
\midrule
\textbf{Overall} & & -0.01657 & -2.4\% \\
\bottomrule
\end{tabular}
\end{table}


\section{Discussion}\label{sec:discussion}

\subsection{GEO Environment and RL Performance}

We introduced GEO, a Gymnasium-compatible environment for optimal execution that 
integrates calibrated transient impact models with vectorized simulation. GEO's 
design enables direct transfer between backtesting and live deployment, reducing 
the sim-to-real gap inherent in custom execution simulators.

\begin{figure}[!p]
    \centering
    \includegraphics[width=\linewidth]{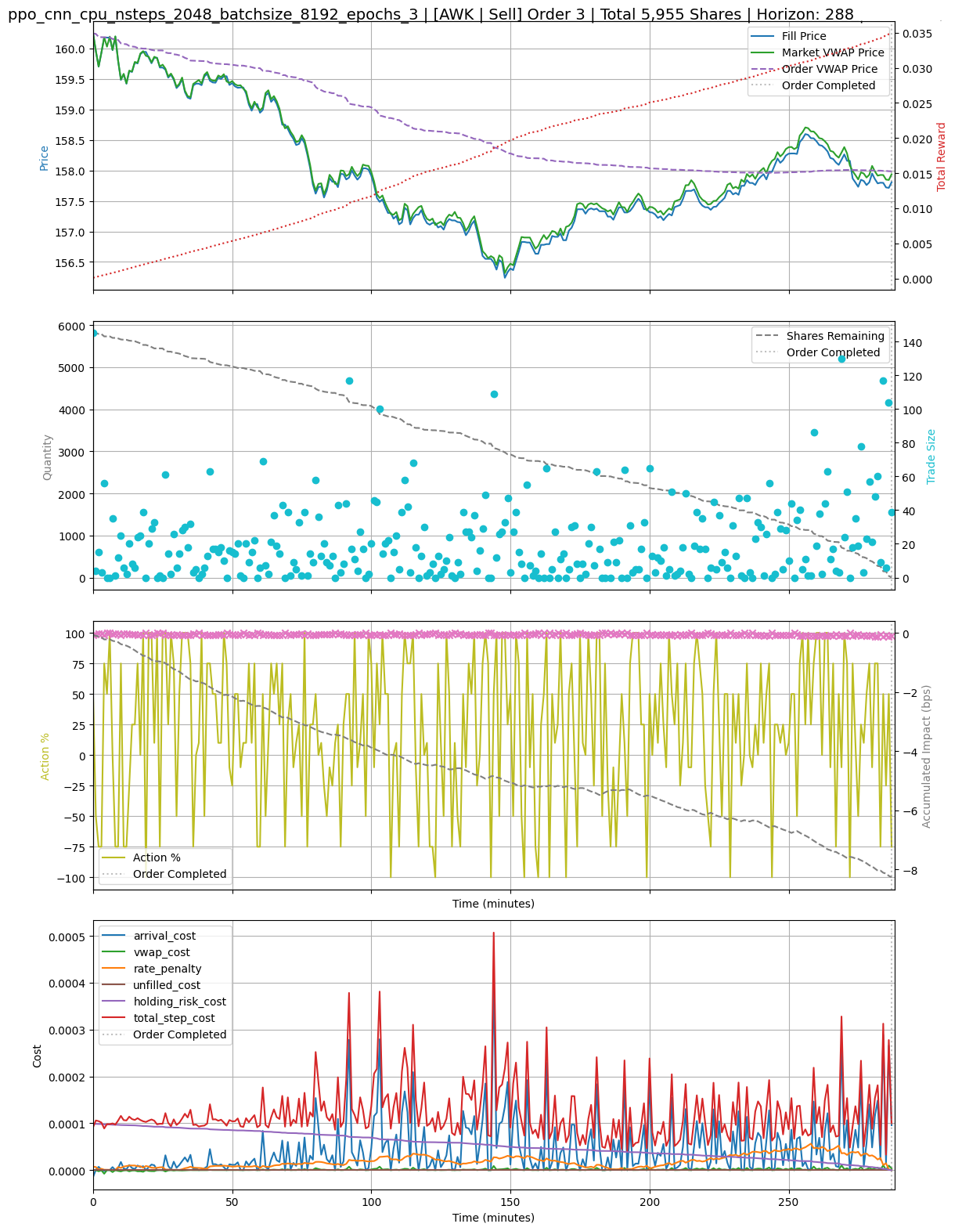}
    \caption{Anatomy of a PPO-CNN RL order showing propagator-driven fill prices, remaining inventory, policy actions, immediate impact, and cost decomposition.}
    \label{fig:order_anatomy}
\end{figure}

Within GEO, PPO-CNN achieved 2.13 bps arrival slippage, outperforming VWAP 
(5.23 bps) and TWAP (7.01 bps) by 59\% and 70\% respectively. Both PPO agents 
reduced total costs to 178 bps, roughly half TWAP's 303 bps, primarily through 
front-loaded schedules that internalize holding costs. 
\edited{Figure~\ref{fig:order_anatomy} shows the ``anatomy'' of a single PPO-CNN order, illustrating how the propagator affects fill prices, how inventory is managed, and how costs accumulate. This integrated view demonstrates how the RL agent perceives state, chooses actions, and experiences costs under the transient impact model.}

The CNN's advantage over MLP (3.78 bps) despite processing identical 13-dimensional 
observations demonstrates that architectural improvements yield measurable gains. 
The convolutional layers enable learning joint patterns between correlated features—price, 
volume, and inventory states—where the MLP treats each independently. Figure~\ref{fig:ppo_trajectories} 
shows the CNN moderating front-loading during adverse price drift, suggesting 
context-dependent adaptation beyond the MLP's static aggressiveness.

These improvements occur within a simulator calibrating exponential impact decay 
on \edited{one-minute level} data with $R^2 \approx$ 0.02-0.10. While this explained variance appears modest, 
it reflects realistic microstructure signal-to-noise ratios. The results represent 
performance under calibrated impact dynamics, not predictions of live execution.

\subsection{Quality-Diversity: Current Results and Future Potential}

MAP-Elites revealed substantial regime heterogeneity. Three cells achieved 8-10\% 
improvements, with high-volatility/medium-liquidity reaching +10.3\%. However, 
high-volatility/low-liquidity degraded -30.2\%, and the overall ensemble averaged 
-2.4\% below baseline.

The pattern is clear: specialists succeed in data-rich regimes with strong signal 
(medium liquidity provides training orders, high volatility amplifies impact 
patterns) but catastrophically overfit in sparse cells. Our implementation used 
simple Gaussian parameter mutations over 500 iterations—a conservative approach 
that prioritizes interpretability over computational efficiency.

Recent advances suggest substantial room for improvement. Parallelized quality-diversity 
\citep{lim2023accelerated} could reduce the 5.5-hour runtime by orders of magnitude. 
Specialized mutation operators for neural networks \citep{faldor2024synergizing} 
may improve exploration efficiency beyond naive Gaussian noise. Methods designed 
for stochastic objectives \citep{flageat2025extractqd} could better handle the 
inherent noisiness of financial data, where fitness evaluation on small order 
samples introduces high variance.

Effective deployment requires validation-based selection—use only specialists 
demonstrating robust out-of-sample gains—and intelligent routing that falls back 
to the baseline in low-confidence regimes. The 3×3 grid may be too coarse; finer 
phenotype partitions or continuous descriptor spaces warrant investigation. With 
these refinements, quality-diversity methods could provide interpretable performance 
maps across market regimes while maintaining robustness.

\subsection{Practical Implications}

RL-based execution appears viable for institutional-scale orders where multi-basis-point 
improvements justify development costs. The CNN architecture provides a strong 
baseline without requiring complex temporal models or extensive feature engineering. 
Quality-diversity methods show promise for discovering regime specialists but 
require substantial compute and careful validation before deployment.

Key limitations remain: our impact model makes stationarity assumptions, calibration 
quality varies across stocks, and all results derive from simulation. Live 
validation would address questions of latency, partial fills, and strategic 
interaction with other market participants. Nevertheless, these results demonstrate 
that RL execution has progressed from academic curiosity toward practical 
consideration for well-resourced trading operations.

\appendix

\newpage

\section{Appendix}\label{sec:appendix}

\subsection{Data Details}

\begin{table}[!h]
\centering
\renewcommand{\arraystretch}{0.9}  
\setlength{\extrarowheight}{-1pt}  

\small  
\caption{Minute bar dataset: Raw and derived data fields by symbol and trading day.}
\label{tab:data_fields}

\rowcolors{2}{gray!10}{white} 
\begin{tabular}{|l|l|l|p{10cm}|}
\hline
\textbf{Field} & \textbf{Symbol} & \textbf{Type} & \textbf{Description / Formula} \\ 
\hline
\texttt{time} & $t$ & Raw & Minute bin within the continuous trading session. \\
\texttt{trade\_count} & $\nu$  & Raw & Number of reported trades in the minute. \\
\texttt{trade\_volume} & $V_t$ & Raw & Total number of shares traded in the minute. \\
\texttt{hid\_vol} & $V^{\mathrm{hidden}}_t$ & Raw & Reported hidden shares traded in the minute. \\
\texttt{unsided\_vol} & $V^{\mathrm{unsided}}_t$ & Raw & Shares traded with unknown aggressor side. \\
\texttt{sell\_vol} & $V^{\mathrm{sell}}_t$ & Raw & Shares traded on the sell side (aggressive seller). \\
\texttt{buy\_vol} & $V^{\mathrm{buy}}_t$ & Raw & Shares traded on the buy side (aggressive buyer). \\
\texttt{bid\_price} & $p^{\mathrm{bid}}_t$ & Raw & Best bid quote price at the end of the minute. \\
\texttt{ask\_price} & $p^{\mathrm{ask}}_t$ & Raw & Best ask quote price at the end of the minute. \\
\texttt{mid\_price} & $m_t$ & Derived & Mid-quote price: $m_t = \frac{p^{\mathrm{bid}}_t + p^{\mathrm{ask}}_t}{2}$. \\
\texttt{bid\_size} & $\delta_{\mathrm{bid}}$ & Raw & Displayed bid size (shares) at the end of the minute. \\
\texttt{ask\_size} & $\delta_{\mathrm{ask}}$ & Raw & Displayed ask size (shares) at the end of the minute. \\
\texttt{trade\_first} & $P_{\mathrm{first}}$ & Raw & First trade price in the minute (removed as predominantly missing). \\
\texttt{trade\_last} & $P_{\mathrm{last}}$ & Raw & Last trade price in the minute. \\
\texttt{trade\_high} & $P_{\mathrm{high}}$  & Raw & Highest trade price in the minute. \\
\texttt{trade\_low} & $P_{\mathrm{low}}$ & Raw & Lowest trade price in the minute. \\
\texttt{vwap} & $P_{\mathrm{vwap}}$  & Raw & Volume-weighted average trade price in the minute. \\
\texttt{trade\_imbalance} & $\epsilon_t$ & Derived & Signed volume imbalance: $\epsilon_t = \frac{V^{\mathrm{buy}}_t - V^{\mathrm{sell}}_t}{V_t}$. \\
\texttt{volatility} &  $\sigma$  & Derived & Realised volatility from a rolling window (default: 21-min rolling standard deviation of mid-price returns). \\
\hline
\end{tabular}
\end{table}

\begin{figure}[H]
  \centering
  \includegraphics[width=0.8\textwidth]{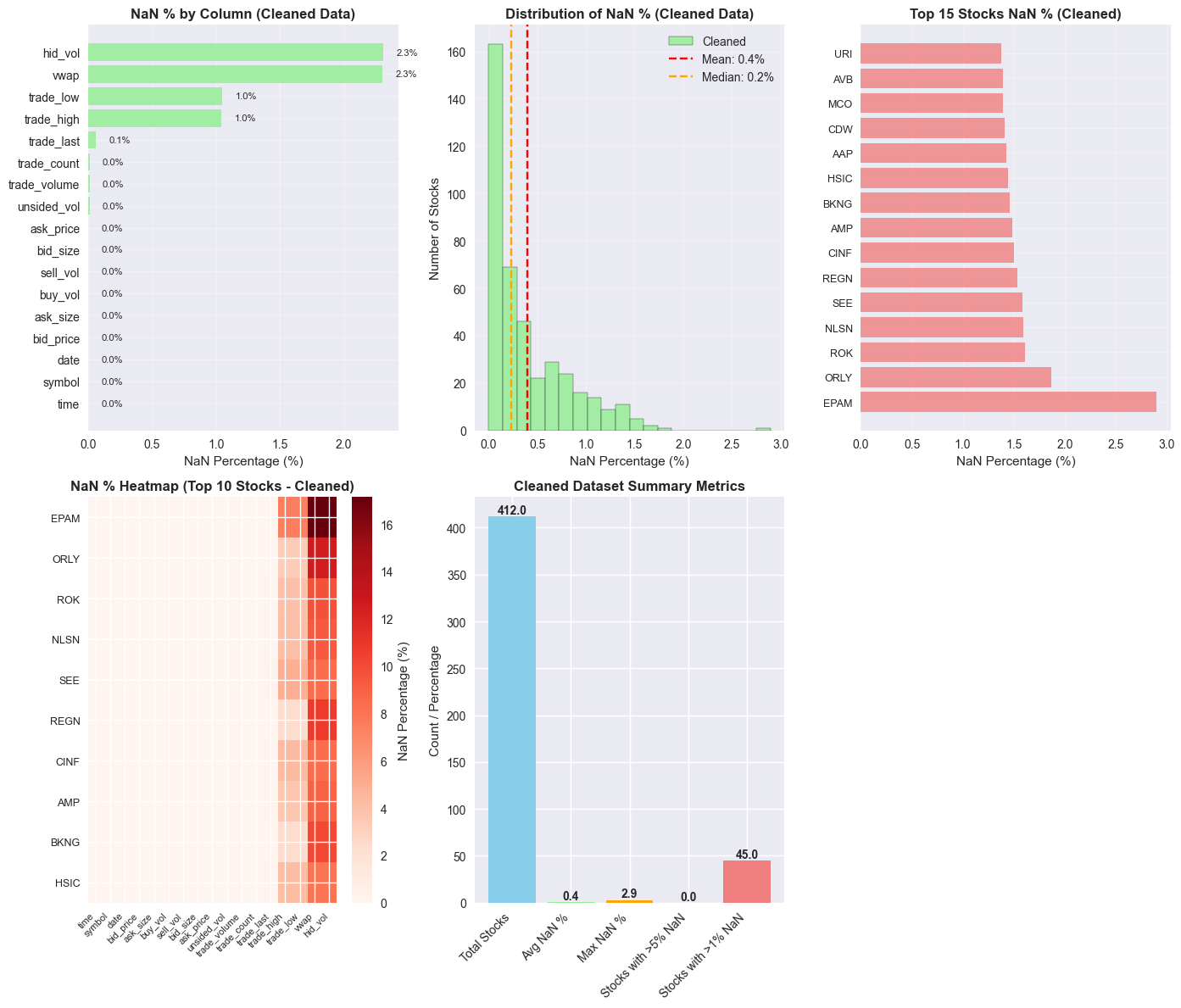}
  \caption{Summary statistics of \edited{cleaned} data.}
  \label{fig:clean_data}
\end{figure}

\setlength{\LTcapwidth}{\textwidth} 

\renewcommand{\arraystretch}{1.3}

\begin{longtable}{|l|l|p{10.5cm}|}
\caption{Daily dataset fields returned by the aggregation pipeline (stored output).}\label{tab:daily_features_returned}\\
\hline
\textbf{Field} & \textbf{Symbol} & \textbf{Description / Formula} \\
\hline
\endfirsthead

\hline
\multicolumn{3}{|l|}{\tablename\ \thetable{} -- \textit{continued from previous page}}\\
\hline
\textbf{Field} & \textbf{Symbol} & \textbf{Description / Formula} \\
\hline
\endhead

\hline
\multicolumn{3}{|r|}{\textit{Continued on next page}}\\
\hline
\endfoot

\hline
\endlastfoot

\rowcolors{2}{gray!10}{white}

\texttt{symbol} & $S$ & Ticker identifier. \\
\texttt{date} & $d$ & Trading day (YYYYMMDD). \\
\hline
\texttt{adv\_21} & $\mathrm{ADV}_{21,d}$ & 21-day rolling average daily volume:
\begin{equation*}
\mathrm{ADV}_{21,d} = \tfrac{1}{21}\sum_{i=0}^{20} V_{d-i}
\end{equation*} \\
\texttt{avg\_trade\_count\_21} & $\overline{\nu}_{21,d}$ & 21-day rolling average of daily trade counts:
\begin{equation*}
\overline{\nu}_{21,d} = \tfrac{1}{21}\sum_{i=0}^{20} \nu_{d-i}
\end{equation*} \\
\texttt{avg\_spread\_21} & $\overline{s}_{21,d}$ & 21-day rolling average of the day-level spread metric (units consistent with input \texttt{spread}). \\
\texttt{avg\_depth\_21} & $\overline{\delta}_{21,d}$ & 21-day rolling average of daily depth, where
\begin{equation*}
\delta_d = \tfrac{\text{bid\_size}_d + \text{ask\_size}_d}{2}, \quad
\overline{\delta}_{21,d} = \tfrac{1}{21}\sum_{i=0}^{20} \delta_{d-i}
\end{equation*} \\
\texttt{vwap} & $P^{\mathrm{VWAP}}_d{}^{\ast}$ & Daily VWAP with fallback: if VWAP is missing, set
$P^{\mathrm{VWAP}}_d{}^{\ast} \leftarrow P^{\mathrm{last}}_d$ (last trade price). \\
\hline
\texttt{daily\_volatility} & $\widehat{\sigma}^{(1)}_d$ & Parkinson volatility (window $w{=}1$ day), clipped to $[10^{-4},2.0]$:
\begin{equation*}
\widehat{\sigma}^{(w)}_d = \sqrt{\tfrac{1}{4w\ln 2}\sum_{i=0}^{w-1}\left[\ln\!\left(\tfrac{P^H_{d-i}}{P^L_{d-i}}\right)\right]^2}
\end{equation*} \\
\texttt{daily\_vol\_lag1} & $\widehat{\sigma}^{(2)}_d$ & Parkinson volatility (window $w{=}2$ days), clipped to $[10^{-4},2.0]$ (same formula as above with $w=2$). \\
\texttt{daily\_vol\_5d} & $\widehat{\sigma}^{(5)}_d$ & Parkinson volatility (window $w{=}5$ days), clipped to $[10^{-4},2.0]$ (same formula as above with $w=5$). \\
\hline
\texttt{trade\_high} & $P^H_d$ & Highest trade price of day $d$. \\
\texttt{trade\_low} & $P^L_d$ & Lowest trade price of day $d$. \\
\hline
\end{longtable}

\subsection{Full list of performance Metrics}\label{subsec:full_perf_metrics}



\begin{longtable}{|p{5.5cm}|p{10cm}|}
\caption{Evaluation metrics used to assess execution policy performance.}
\label{tab:eval_metrics} \\
\hline
\textbf{Metric} & \textbf{Mathematical Definition} \\
\hline
\endfirsthead

\hline
\multicolumn{2}{|l|}{\tablename\ \thetable{} -- \textit{continued from previous page}}\\
\hline
\textbf{Metric} & \textbf{Mathematical Definition} \\
\hline
\endhead

\hline
\multicolumn{2}{|r|}{\textit{Continued on next page}}\\
\hline
\endfoot

\hline
\endlastfoot

\hline
\textbf{Arrival price slippage} &
\begin{equation*}
C_{\mathrm{arrival}} = 10^4 \cdot \mathrm{side}\left(
\frac{\sum_{t=0}^{H-1} p^{\mathrm{fill}}_t\, q_t}{Q_0} - p_0
\right)\Big/ p_0
\end{equation*} \\
\hline
\multicolumn{2}{|p{16cm}|}{Implementation shortfall vs.\ arrival mid-price $p_0$, in basis points.} \\
\hline

\textbf{Market VWAP vs.\ Arrival} &
\begin{equation*}
C_{\mathrm{mktVWAP}} = 10^4 \cdot \left(
\frac{\sum_{t=0}^{H-1} p^{\mathrm{mkt}}_t\, V_t}{\sum_{t=0}^{H-1} V_t} - p_0
\right)\Big/ p_0
\end{equation*} \\
\hline
\multicolumn{2}{|p{16cm}|}{Reference measure of how the market VWAP itself moved relative to arrival, isolating market drift.} \\
\hline

\textbf{VWAP slippage} &
\begin{equation*}
C_{\mathrm{VWAP}} = \mathrm{side}\left(
\frac{\sum_{t=0}^{H-1} (p^{\mathrm{fill}}_t)\, q_t}{Q_0} - P^{\mathrm{VWAP}}
\right)
\end{equation*} \\
\hline
\multicolumn{2}{|p{16cm}|}{Performance relative to the VWAP benchmark} \\
\hline

\textbf{Completion rate} &
\begin{equation*}
\frac{\sum_{t=0}^{H-1} q_t}{Q_0}
\end{equation*} \\
\hline
\multicolumn{2}{|p{16cm}|}{Proportion of shares executed by horizon $H$; target is 1 (100\%).} \\
\hline

\textbf{Horizon usage} &
\begin{equation*}
\frac{t_{\mathrm{last\_trade}}}{H}
\end{equation*} \\
\hline
\multicolumn{2}{|p{16cm}|}{Fraction of horizon consumed before order completion; lower values imply earlier execution.} \\
\hline

\textbf{Action variability} &
\begin{equation*}
\mathrm{Var}\!\big(a_{0:H-1}\big)
\end{equation*} \\
\hline
\multicolumn{2}{|p{16cm}|}{Variance of policy actions; high variability may indicate unstable strategies.} \\
\hline

\textbf{No-trade percentage} &
\begin{equation*}
\frac{\#\{\,t:\, q_t = 0\,\}}{H}
\end{equation*} \\
\hline
\multicolumn{2}{|p{16cm}|}{Fraction of minutes where no trades were executed; indicates inactivity.} \\
\hline

\textbf{High-rate in favourable periods} &
\begin{equation*}
\frac{\#\{\,t:\, q_t > \rho_{\mathrm{target}} V_t \ \wedge\ p^{\mathrm{stock}}_t < P^{\mathrm{VWAP}}_t\,\}}{H}
\end{equation*} \\
\hline
\multicolumn{2}{|p{16cm}|}{Share of minutes where the agent accelerated trading when the stock outperformed market VWAP.} \\
\hline

\textbf{Low-rate in unfavourable periods} &
\begin{equation*}
\frac{\#\{\,t:\, q_t < \rho_{\mathrm{target}} V_t \ \wedge\ p^{\mathrm{stock}}_t > P^{\mathrm{VWAP}}_t\,\}}{H}
\end{equation*} \\
\hline
\multicolumn{2}{|p{16cm}|}{Share of minutes where the agent slowed trading when the stock underperformed market VWAP.} \\
\hline

\end{longtable}

\newpage

\bibliographystyle{apalike}
\bibliography{refs}

\begin{thebibliography}{}

\bibitem[Almgren and Chriss, 2001]{AlmgrenChriss2001JOR}
Almgren, R. and Chriss, N. (2001).
\newblock Optimal execution of portfolio transactions.
\newblock {\em The Journal of Risk}, 3(2):5--40.

\bibitem[Amrouni et~al., 2022]{amrouni2021abidesgym}
Amrouni, S., Moulin, A., Vann, J., Vyetrenko, S., Balch, T., and Veloso, M.
  (2022).
\newblock {ABIDES}-gym: gym environments for multi-agent discrete event
  simulation and application to financial markets.
\newblock In {\em Proceedings of the Second ACM International Conference on AI
  in Finance}, ICAIF '21, New York, NY, USA. Association for Computing
  Machinery.

\bibitem[Bouchaud, 2010]{Bouchaud2010PriceImpact}
Bouchaud, J.-P. (2010).
\newblock Price impact.
\newblock In {\em Encyclopedia of Quantitative Finance}. John Wiley \& Sons.

\bibitem[Bouchaud et~al., 2018]{bouchaud2018propagator}
Bouchaud, J.-P., Bonart, J., Donier, J., and Gould, M. (2018).
\newblock The propagator model.
\newblock In {\em Trades, Quotes and Prices: Financial Markets Under the
  Microscope}, pages 252--253. Cambridge University Press.

\bibitem[Bucci et~al., 2019]{Bucci2019crossover}
Bucci, F., Benzaquen, M., Lillo, F., and Bouchaud, J.-P. (2019).
\newblock Crossover from linear to square-root market impact.
\newblock {\em Physical Review Letters}, 122:108302.

\bibitem[Chatzilygeroudis et~al., 2021]{Chatzilygeroudis2021}
Chatzilygeroudis, K., Cully, A., Vassiliades, V., and Mouret, J.-B. (2021).
\newblock Quality-diversity optimization: A novel branch of stochastic
  optimization.
\newblock In Pardalos, P.~M., Rasskazova, V., and Vrahatis, M.~N., editors,
  {\em Black Box Optimization, Machine Learning, and No-Free Lunch Theorems},
  pages 109--135. Springer, Cham.

\bibitem[Cont et~al., 2014]{Cont2010impact}
Cont, R., Kukanov, A., and Stoikov, S. (2014).
\newblock The price impact of order book events.
\newblock {\em Journal of Financial Econometrics}, 12:47--88.

\bibitem[Faldor et~al., 2025]{faldor2024synergizing}
Faldor, M., Chalumeau, F., Flageat, M., and Cully, A. (2025).
\newblock Synergizing quality-diversity with descriptor-conditioned
  reinforcement learning.
\newblock {\em ACM Transactions on Evolutionary Learning and Optimization},
  5(1):3 (35 pages).

\bibitem[Flageat et~al., 2025]{flageat2025extractqd}
Flageat, M., Huber, J., Helenon, F., Doncieux, S., and Cully, A. (2025).
\newblock Extract-{QD} framework: A generic approach for quality-diversity in
  noisy, stochastic or uncertain domains.
\newblock In {\em Proceedings of the Genetic and Evolutionary Computation
  Conference}, GECCO '25, pages 140--148, New York, NY, USA. Association for
  Computing Machinery.

\bibitem[Gatheral et~al., 2012]{gatheral2012transient}
Gatheral, J., Schied, A., and Slynko, A. (2012).
\newblock Transient linear price impact and {F}redholm integral equations.
\newblock {\em Mathematical Finance}, 22(3):445--474.

\bibitem[Hafsi and Vittori, 2024]{hafsi2024optimal}
Hafsi, Y. and Vittori, E. (2024).
\newblock Optimal execution with reinforcement learning.
\newblock {\em arXiv preprint arXiv:2411.06389}.

\bibitem[Hendricks and Wilcox, 2014]{hendricks2014reinforcement}
Hendricks, D. and Wilcox, D. (2014).
\newblock A reinforcement learning extension to the {A}lmgren-{C}hriss
  framework for optimal trade execution.
\newblock In {\em 2014 IEEE Conference on Computational Intelligence for
  Financial Engineering \& Economics (CIFEr)}, pages 457--464.

\bibitem[Jerome et~al., 2023]{jerome2023mbtgym}
Jerome, J., S\'{a}nchez-Betancourt, L., Savani, R., and Herdegen, M. (2023).
\newblock Mbt-gym: Reinforcement learning for model-based limit order book
  trading.
\newblock In {\em Proceedings of the Fourth ACM International Conference on AI
  in Finance}, ICAIF '23, page 619–627, New York, NY, USA. Association for
  Computing Machinery.

\bibitem[Lim et~al., 2023]{lim2023accelerated}
Lim, B., Allard, M., Grillotti, L., and Cully, A. (2023).
\newblock Accelerated quality-diversity through massive parallelism.
\newblock {\em Transactions on Machine Learning Research}.

\bibitem[{Mana Tech LLC}, 2025]{manatechllc}
{Mana Tech LLC} (2025).
\newblock Historical u.s. equities market data.
\newblock \url{https://manatech.ai/}.
\newblock Accessed: 2025-08-17.

\bibitem[Mastromatteo et~al., 2014]{mastromatteo2014agent}
Mastromatteo, I., T{\'o}th, B., and Bouchaud, J.-P. (2014).
\newblock Agent-based models for latent liquidity and concave price impact.
\newblock {\em Physical Review Letters}, 113(26):268701.

\bibitem[Mouret and Clune, 2015]{mouret2015mapelites}
Mouret, J.-B. and Clune, J. (2015).
\newblock Illuminating search spaces by mapping elites.
\newblock {\em arXiv preprint arXiv:1504.04909}.

\bibitem[Nevmyvaka et~al., 2006]{nevmyvaka2006reinforcement}
Nevmyvaka, Y., Feng, Y., and Kearns, M.~J. (2006).
\newblock Reinforcement learning for optimized trade execution.
\newblock In {\em Proceedings of the 23rd International Conference on Machine
  Learning (ICML)}, pages 673--680. ACM.

\bibitem[Obizhaeva and Wang, 2013]{ObizhaevaWang2013}
Obizhaeva, A.~A. and Wang, J. (2013).
\newblock Optimal trading strategy and supply/demand dynamics.
\newblock {\em Journal of Financial Markets}, 16(1):1--32.

\bibitem[Parkinson, 1980]{parkinson1980extreme}
Parkinson, M. (1980).
\newblock The extreme value method for estimating the variance of the rate of
  return.
\newblock {\em Journal of Business}, 53(1):61--65.

\bibitem[Perold, 1988]{Perold1988IS}
Perold, A.~F. (1988).
\newblock The implementation shortfall: Paper vs. reality.
\newblock {\em The Journal of Portfolio Management}, 14(3):4--9.

\bibitem[{Ray Team}, 2025]{RLlibGymnasium}
{Ray Team} (2025).
\newblock {RLlib} environments: Farama {G}ymnasium.
\newblock \url{https://docs.ray.io/en/latest/rllib/rllib-env.html}.
\newblock Accessed: 2026-01-29.

\bibitem[Schulman et~al., 2016]{schulman2016gae}
Schulman, J., Moritz, P., Levine, S., Jordan, M., and Abbeel, P. (2016).
\newblock High-dimensional continuous control using generalized advantage
  estimation.
\newblock In {\em Proceedings of the International Conference on Learning
  Representations (ICLR)}.

\bibitem[Schulman et~al., 2017]{schulman2017ppo}
Schulman, J., Wolski, F., Dhariwal, P., Radford, A., and Klimov, O. (2017).
\newblock Proximal policy optimization algorithms.
\newblock {\em arXiv preprint arXiv:1707.06347}.

\bibitem[{Stable-Baselines3 Developers}, 2025]{SB3Gymnasium}
{Stable-Baselines3 Developers} (2025).
\newblock Using custom environments ({G}ymnasium interface).
\newblock
  \url{https://stable-baselines3.readthedocs.io/en/master/guide/custom_env.html}.
\newblock Accessed: 2026-01-29.

\bibitem[Sutton and Barto, 2018]{sutton2018reinforcement}
Sutton, R.~S. and Barto, A.~G. (2018).
\newblock {\em Reinforcement Learning: An Introduction}.
\newblock MIT Press, 2nd edition.

\bibitem[Towers et~al., 2024]{towers2024gymnasium}
Towers, M., Kwiatkowski, A., Terry, J., Balis, J.~U., De~Cola, G., Deleu, T.,
  Goulão, M., Kallinteris, A., Krimmel, M., Arjun, K.~G., Perez-Vicente, R.,
  Pierré, A., Schulhoff, S., Tai, J.~J., Tan, H., and Younis, O.~G. (2024).
\newblock Gymnasium: A standard interface for reinforcement learning
  environments.
\newblock {\em arXiv preprint arXiv:2407.17032}.

\bibitem[Tóth et~al., 2011]{toth2011anomalous}
Tóth, B., Lempérière, Y., Deremble, C., de~Lataillade, J., Kockelkoren, J.,
  and Bouchaud, J.-P. (2011).
\newblock Anomalous price impact and the critical nature of liquidity in
  financial markets.
\newblock {\em Physical Review X}, 1(2):021006.

\end{thebibliography}

\clearpage

\end{document}